\documentclass[12pt]{iopart} 
\usepackage{iopams} 
\usepackage{cite}
\usepackage{epsfig}
\usepackage{caption}
\usepackage{psfig}
\usepackage{amssymb,bbm}

\newcommand{\Rt}{\mbox{Re}}

\newcommand{\ket}[1]{\, | #1 \rangle}

\newcommand{\Hb}{{\hat{\mathcal H}}_{\beta}}

\newcommand{\ep}{\epsilon}
\newcommand{\ps}{\ps}
\newcommand{\dk}{$\delta$-kicked rotor}
\newcommand{\Uop}{{\hat U}}

\newcommand{\Hop}{{\hat H}}
\newcommand{\Ih}{{\hat I}}
\newcommand{\En} {{\mathcal E}}

\newcommand{\Up}{{\hat{\mathcal U}}} 
 
\newcommand{\Rp}{{\hat{\mathcal R}}} 
 
\newcommand{\Np}{{\hat{\mathcal N}}}

\newcommand{\Ub}{\Up_{\beta}}

\newcommand{\pSE}{{p_{\mbox{\tiny SE}}}}
\newcommand{\qSE}{{q_{\mbox{\tiny SE}}}}
\newcommand{\qSEt}{{q^t_{\mbox{\tiny SE}}}}
\newcommand{\nSE}{{n_{\mbox{\tiny SE}}}}
\newcommand{\tSE}{{\tau_{\mbox{\tiny SE}}}}
\newcommand{\epsm}{$\epsilon$SM}

\newcommand{\xise}{{\tilde\xi}}
\newcommand{\deltot}{{\tilde\delta}_t}

\begin{document} 
 
\title[QR]{Quantum resonances and  
decoherence for $\delta-$kicked atoms} 
 
\author{Sandro Wimberger$^{1,2,3}$, Italo~Guarneri$^{2,3,4}$ and  
Shmuel~Fishman$^{5}$}  
 
\address{$^{1}$Max-Planck-Institut f\"ur Physik komplexer Systeme, 
N\"othnitzer Str. 38, D-01187 Dresden \\ 
$^{2}$Universit\`a degli Studi dell' Insubria, Via Valleggio 11, 
I-22100 Como\\ 
$^{3}$Istituto Nazionale per la Fisica della Materia, Unit\`a di Milano, 
Via Celoria, I-20133 Milano \\ 
$^{4}$Istituto Nazionale di Fisica Nucleare, Sezione di Pavia, Via Bassi 6, 
hI-27100 Pavia\\ 
$^{5}$Physics Department, Technion, Haifa IL-32000 \\ 
} 
\address{e-mail: saw@mpipks-dresden.mpg.de} 
 
\begin{abstract} 
The quantum resonances occurring with $\delta-$kicked atoms when 
the kicking period is an integer multiple of the half-Talbot time 
are analyzed in detail. Exact results about the momentum distribution at 
exact resonance 
are established, both in the case of totally 
coherent dynamics and in the case when decoherence is induced 
by Spontaneous Emission.  A description of the dynamics 
when the kicking period is close to, but not exactly at resonance, is 
derived by means of a quasi-classical approximation where the detuning 
from exact resonance plays the role of the Planck constant. In this way 
scaling laws describing the shape of the resonant peaks are obtained. 
Such analytical results are supported by extensive numerical simulations,   
and explain  some recent surprising experimental observations.   
\end{abstract} 
 

\section{Introduction.}

The present work is devoted to a detailed theoretical 
analysis of some of the quantum resonances occurring in the \dk, 
motivated by laboratory results obtained with experimental 
realizations of that system.

\subsection{Background.}
\label{back1}

The \dk~is a paradigm model in quantum chaology,  
and its theoretical importance 
is connected with the long-time properties of its evolution. 
Among these are effects of 
purely quantum origin such as dynamical localization 
\cite{CCFI79,FGP82,CS8nn,CGS88,GSMKR88,BCGS89,ABMW91,Fis93} and 
the quantum resonances \cite{IS80,Isr90}. Despite 
compelling numerical evidence, dynamical localization 
could not be mathematically proven until very recently\footnote{A 
proof for the case of sufficiently small kicking strength  has been 
announced by J.~Bourgain and S.~Jitomirskaya 
while the present paper was being prepared for publication.}. 
The origin and the main dynamical and spectral features of the 
quantum resonances are instead well understood in general 
mathematical terms \cite{IS80,CG84,Isr90,BB91}. They  
 are observed when the kicking frequency is commensurate to the 
natural frequencies of the rotor and typically result in a
quadratic growth of the energy asymptotically in time. 
\\
Although the \dk~is an abstract theoretical model, it has found an 
experimental realization using tools of atom optics, by a technique pioneered 
by Raizen and coworkers \cite{MRBWR94}. 
In present-day experiments laser-cooled Cesium atoms are 
periodically driven with a standing electromagnetic wave. 
As the frequency of the wave is  slightly detuned 
from an internal atomic transition, a net force on the centres  
of mass of the atoms arises, proportional to the square of the driving 
field \cite{CDG92,GSZ92}.
The standing wave is periodic in space, so the atoms are subject to a 
periodic potential. The latter is turned on and off periodically in time, 
resulting in a sequence of short pulses, or ``kicks''. 
In such experiments, the kicked atoms behave as individual particles 
with negligible interaction due to collision etc. 
\cite{ABGRD97,KOSR99}, so a single-particle theory is applicable.
The nonzero duration of the kicking pulses  
sets a bound on the momentum range wherein the ideal \dk~model is 
applicable. At large momenta the driving becomes adiabatic, leading to  
trivial localization in momentum for the classical and the quantum version of 
the rotor \cite{BFS86,KOSR99}. 
Important properties of the \dk~ have been experimentally reproduced: 
{\it e.g.}, the exponential  
localization in momentum \cite{MRBWR94,BRMSNR99} (away from resonances).   
In addition the influence of noise and decoherence 
\cite{AGSC98,KOSR98,SMOR00,MSOR00},   
and the effect of gravity in the case when the 
kicking direction is vertical 
have been examined in experiments (with possible applications in atomic  
interferometry \cite{GDOSB00,FGR02b,FGR02} and quantum measurement theory 
\cite{Dar02,SDGCGS02,DGCS02new,WGB03}).\\ 
While dynamical localization is a robust phenomenon, the quantum 
resonances are rather sensitive instead, and the correspondence 
between theory and experiments  is less appealing in their case.  
 The following items represent  
incomplete or missing matches between the \dk~ theory 
and experimental results \cite{OSMKR00,DGOCS01,DGOSBG01}:\\ 
(i) quantum resonances occur for the \dk~ whenever the kicking 
period $\tau$ (in appropriate units) is a rational multiple of $4\pi$, which
is the Talbot time in the language of diffraction optics
\cite{DGOCS01,DB1997,Goo1996}. In
experiments, only at the values $\tau =2\pi, 4\pi, 6\pi$, a special 
behaviour has been observed, in the form of narrow peaks in the energy 
absorption vs. $\tau$. This means that higher order resonances have not been 
resolvable within the experimental bounds. \\ 
(ii) {\it no quadratic} energy growth at resonance has been experimentally 
observed. "Ballistic peaks" were reported by Raizen {\it et al.} 
\cite{OSMKR00}, due to a tiny fraction of atoms which indeed show a 
quadratically increasing energy. This is in contrast to the bulk of the atomic 
cloud, which is instead frozen in a rather narrow distribution in 
momentum. The latter is, however, not exponential, at variance with the 
distribution found at non-resonant values of $\tau$, i.e.~ in the localized 
regime. \\ 
(iii) {\it an enhancement} of the resonant peaks has 
been experimentally observed 
with {\it increasing} degree of decoherence. 
This is especially
surprising, because decoherence is expected to drive the 
quantum system towards a more classical behaviour,
hence to inhibit purely quantum effects, 
such as the quantum resonances of the \dk. 
As to dynamical localization, its destruction  due to decoherence 
was experimentally observed in
\cite{BBGSSW91,AGSC98,KOSR98}\footnote{Similarly noise due to
Spontaneous Emission affects non-dispersive quantum 
wave packets formed in highly excited Rydberg atoms \cite{HB98,BDZ02}.}.
\\  
The listed findings do not match with the theory of the \dk, 
and various physical reasons have been identified. 
The simplest, yet 
deepest of these is the starting point of the present paper: the experiments 
do not provide a 
realization of the  \dk~ but of the $\delta$-kicked particle instead, the 
difference being that while rotors move in circles,  particles move 
in (approximately) straight lines. 
This difference is irrelevant 
on the classical level, however it has nontrivial quantum 
implications, that lie within the scope of the Bloch theory.
The quantum dynamical localization  is not 
impaired (localization lengths and fluctuation properties are affected, 
though)\cite{Isr90}. Not so in what concerns the quantum resonances, that make 
the subject of this work.  

\subsection{Outline, and statement of results.}

In this paper we analyze in  
mathematical detail the dynamics of atoms at quantum resonance 
and in its vicinity  
for  the ideal case of $\delta$-kicks. Our basic tool is the Bloch theory 
for quantum particles in periodic potentials. The basic 
constructions  of that theory 
as applied to $\delta-$kicked particles are shortly reviewed in 
section~\ref{bloch2}. Our main results are:

\begin{enumerate}

\item  In the absence of decoherence, we show that  the  
mean energy increases linearly 
with time (section~\ref{avereng}) yet the motion is dynamically localized,  
in the sense that 
the momentum distribution settles in the course of time to a relatively narrow 
steady state distribution (section~\ref{coarsedis}). 
This  seeming contradiction is 
resolved by the slow algebraic decay of  the tails 
of the steady state distribution. 
We give  an integral expression 
(eq.~(\ref{fourdis}),(\ref{cgdis}),(\ref{pstat1})) 
of the latter distribution, whence we derive 
 exact bounds  (\ref{estdist}) and large-momentum asymptotics (\ref{decdis}).

\item We derive a scaling law (section~\ref{dynres}) 
 that describes the shape 
of the resonant peaks which are observed in the energy vs. kicking 
period curves. The scaling  law (\ref{repla0}) 
is then demonstrated  by numerical simulations. It 
is derived by means of a quasi-classical approximation  
(subsection~\ref{epsclass4}), where the role of the Planck constant is  
played by the detuning from exact resonance. 
This technique was introduced in \cite{FGR02b,FGR02}.  
It allows to describe the near-to-resonant 
quantum motion in terms of a Standard Map, which is different 
in parameter values from the one that is obtained in the usual
classical limit of the \dk. The {\it quantum} resonance exactly corresponds 
to the main {\it classical} resonance of this standard map.  
The stable island associated with the latter resonance 
accounts for 
the structure of the resonant peaks. This analysis 
closely links quantum resonances 
with classical nonlinear ones, thus closing a gap between different meanings 
of the term ``resonance'', which were deemed to be unrelated. 

\item  We devise a model of stochastic \dk~dynamics, where the 
external noise is modelled after 
the Spontaneous Emission (SE) effects that are used  to induce controlled 
decoherence in experiments (section~\ref{spontanem5}). 
Our statistical assumptions (subsection~\ref{model}) are a 
compromise between physical adherence and mathematical convenience, 
and allow to derive simple exact results, that expose the 
key mechanism whereby the quantum resonances are affected. 
We explicitely compute the growth of the mean energy in time 
(eq.~(\ref{engyfin})). 
We prove  that no steady state  momentum distribution is any more attained, 
and that, in contrast, the momentum distribution evolves into  
a Gaussian distribution which spreads diffusively (subsection~\ref{assym}).  
By checking  such results against realistic numerical 
simulations we demonstrate  that their validity does not crucially 
depend on our technical assumptions (subsection~\ref{disnumr}). 
So they  are relevant for a wide range of experimental situations. 

\item The stochastic \dk~dynamics provides a remarkable  example where 
the classical and quantum effects of the coupling to the environment can be 
separated. This is achieved by a stochastic gauge transformation 
(section~\ref{modsse}). 

\item We provide a 
description of the nearly resonant dynamics in the presence of 
decoherence (section~\ref{epswithnoise}), by means of the small-detuning 
quasi-classical  asymptotics (point (ii) above). 
In this way we derive a scaling law 
for the structure of the 
resonant peaks in the presence of decoherence (eq.~(\ref{replace})), 
which is then demonstrated  by numerical simulations.

\item  We explain  the experimentally observed enhancement 
of the resonant peaks in the presence of SE (section~\ref{exp6}). 
At a low degree of decoherence, the 
linear growth of energy at resonance (eq.~(\ref{engyfin})) 
is but slightly changed with 
respect to the decoherence-free case,   
so  decoherence cannot by 
itself explain the much stronger enhancement of the resonant 
peaks that was experimentally observed. In order 
to understand the latter we analyse   
differences which experiments inevitably impose with respect to the 
ideal models. 
Experimental cutoffs in momentum exist,  due on one hand to the 
finite duration of the pulses and on the other hand to the signal-to-noise 
ratio. Using the results of point (i) we argue, 
and numerically demonstrate, that  the latter type 
of cutoff is responsible for the observed enhancement. 
The coherent energy growth at resonance is due only to the fastest atoms, 
hence it is stopped as soon as such atoms escape the finite bounds of 
experimental observation. This peculiarity is rapidly destroyed 
by decoherence, which calls for all the atoms to participate in the 
mean energy growth.
 
\end{enumerate}

\section{Kicked atoms vs. Kicked Rotors.} 
\label{bloch2} 
 
We consider a one-dimensional model for a kicked atom of mass $M$, subject to  
time- and space-periodic kicks, with periods $T$ and $\pi k_{L}^{-1}$, 
respectively. $k_{L}$ is the wave number of the kicking field 
whose maximum  
strength is given by $\kappa$. We rescale momentum in units 
of $2\hbar k_{L}$,  position in units of $(2k_{L})^{-1}$, 
mass in units of $M$. Energy is then given in units of $\hbar^2 
(2k_{L})^2/M$, time in units of $M/\hbar (2k_{L})^2$, and the 
reduced Planck's constant equals 1. The dynamics of the kicked atoms is
induced by the following Hamiltonian \cite{FGR02,FGR02b}, 
which depends on the continuous time parameter $t'$: 
\begin{equation}
\Hop (t') =\frac{\hat{P}^2}{2} + k\cos(\hat{X})\sum_{t=-\infty}^{+\infty}
 \delta (t'-t\tau)\;,
\label{ham}
\end{equation}
where $\hat{X},\hat{P}$ are the position and the momentum operator  
respectively, 
$k=\kappa/\hbar$, and the kicking period $\tau =\hbar T(2k_{L})^2/M$. 
The state evolution of the atom from one 
kick to immediately after the next kick is determined by the unitary 
Floquet operator:  
\begin{equation} 
\label{floquet} 
\Uop= e^{-ik\cos(\hat{X})}e^{-i\tau \hat{P}^2/2}.
\end{equation} 
Iterated application of the one-cycle operator $\Uop$  
yields the dynamics of the 
atom in the discrete time given by the kick counter, which we 
denote by the {\it integer} $t$. We further denote  
$\ket {\psi}$ the state vector of the atom, and $\psi (x) =
\langle x \ket {\psi}$, ${\tilde\psi}(p)=\langle p|\psi\rangle$ 
the wave functions in the position and in the momentum 
representation, respectively. 
\\
The model (\ref{floquet}) 
differs from the kicked-rotor model because the 
particle does not move on a circle but on a line instead.    
A link between the two models is generated by the spatial  
periodicity of the kicking potential: the Floquet operator (\ref{floquet})  
commutes with 
spatial translations by multiples of $2\pi$. As is well-known from the 
Bloch theory, this enforces conservation  
of the Quasi-Momentum (QM). In our units, QM is given by the fractional part 
$\{p\}$ of the 
momentum $p$ and will be denoted   
$\beta,\;(0\leq \beta <1)$. For a sharply defined value of 
quasi-momentum, the wave function of the atom is 
a Bloch wave, of the form $\exp(i\beta x) \psi_{\beta}(x)$, with    
$\psi_{\beta}(x)$ a $2\pi-$periodic function. The general particle 
wave packet is  obtained by superposing Bloch waves parametrized by the 
continuous variable $\beta\in[0,1)$:  
\begin{equation} 
\label{recx} 
\psi (x)=\int_0^1d\beta\;e^{i\beta x}\psi_{\beta}(x). 
\end{equation} 
Denoting $\theta \equiv x$ mod$(2\pi)$, we in turn have 
\begin{equation} 
\label{psibeta} 
\psi_{\beta}(\theta)= \frac{1}{\sqrt{2\pi}} 
\sum\limits_{n} {\tilde \psi} (n+\beta)\; e^{in\theta}. 
\end{equation}
In the special case when 
 the state of the particle is a plane wave with  
momentum $p_0$, then  
\begin{equation}
\label{plw}
\psi_{\beta} (\theta)~=~(\sqrt{2\pi})^{-1} \delta (\beta-\beta _0)
\exp(in_0\theta)\;,
\end{equation}   
where 
$\beta _0 =\{p_0\}$ and $n_0=[p_0]$ are the fractional and integer part of 
$p_0$, respectively.
    
For any given $\beta$, $\psi_{\beta}(\theta)$ may be thought of as the  
wave function of a rotor  
with angular coordinate $\theta$, henceforth called $\beta-$rotor. We denote  
the corresponding state of the rotor by $\ket {\psi_{\beta}}$.  
From (\ref{floquet}) and (\ref{psibeta}), it follows that, while $\ket 
{\psi}$ evolves into $\Uop \ket {\psi}$, 
$\ket {\psi_{\beta}}$ evolves into $\Ub \ket {\psi_{\beta}}$, with 
\begin{equation} 
\label{ubeta} 
\Ub \;=\;e^{-ik\cos({\hat \theta})}\;e^{-i\frac{\tau}{2}(\Np+\beta)^2}, 
\end{equation} 
where $\Np$ is the angular momentum operator: in the  
$\theta-$representation, $\Np=-id/d\theta$ with periodic boundary conditions.  
The Floquet operator (\ref{ubeta}) differs from the Floquet 
operator of the standard \dk~ 
by the phase $\beta$. In previous studies of this rotor variant, $\beta$ was 
typically regarded as an external Aharonov-Bohm flux \cite{Isr86}. 
 
\section{Quantum resonances.} 
\label{ensemble3} 
 
The time evolution determined by the Floquet operator (\ref{ubeta}) preserves 
many of the dynamical properties of the standard \dk~ (corresponding 
to $\beta=0$), but not 
necessarily the quantum resonances. These only occur for special values 
of quasi-momentum \cite{Isr90}. Whenever   
$\tau =4\pi p/q$ ($p,q$ mutually prime integers), and 
$\beta =m/2p$ 
with $m$ an integer such that $0\leq m< 2p$,  
the operator (\ref{ubeta}) commutes with translations in 
momentum space by multiples of $q$, leading to bands in its  
quasi-energy spectrum. 
For special values of $q$, the bands may be flat,
i.e., of zero width; 
such is the case, e.g.,   
for $q=2$, $m$ an even integer. 
In typical cases, however,  the bands are not flat and 
result in  absolutely continuous quasi-energy spectra, that 
enforce  ballistic growth of momentum.   
The width of the bands rapidly decreases as the order $q$ of the 
resonance increases \cite{IS80,Isr90}, so  
ballistic motion is observable only after quite long times.  
This makes higher-order  
resonances experimentally hardly detectable, and provides an 
already well-known answer to   
the first problem (i) stated in the Introduction (subsection~\ref{back1}). 
Further reasons hindering experimental observation of higher resonances are
pointed out in section~\ref{exp6}.

\subsection{$\beta -$rotor dynamics.} 
\label{singlebeta} 
 
We focus on the main resonances  
$q=1,2$ in the following, i.e., we set $\tau =2\pi \ell$, with 
$\ell$ a positive integer. Inserting this value  
for the kicking period into (\ref{ubeta}), and using the identity  
$\exp(-i\pi n^2\ell)=\exp(-i\pi n\ell)$,  
we obtain (apart from an $n$-independent phase factor)
\begin{equation} 
\label{ubetapsi} 
\Ub\;=\; e^{-ik\cos({\hat \theta})}\;e^{-i\xi\Np} 
\end{equation} 
where $\xi\equiv \pi \ell (2\beta\pm 1)$mod$(2\pi)$ will be taken 
in $[-\pi,\pi)$.  
The 2nd operator on the rhs 
of (\ref{ubetapsi}) will be denoted $\Rp(\xi)$. In the $\theta-$representation 
it acts according to:
\begin{equation}
\label{Rop}
(\Rp(\xi)\psi_{\beta})(\theta)=\psi_{\beta}(\theta-\xi)\;.
\end{equation}
The state after the $t-$th kick is then given by 
\begin{equation} 
\label{ubetatpsi} 
(\Ub^t {\psi _{\beta}}) (\theta) = e^{-ik F(\theta,\xi,t)} 
\psi_{\beta} (\theta -t\xi)\;, 
\end{equation} 
with
\begin{equation} 
\label{effd} 
F(\theta,\xi,t) = \sum_{s=0}^{t-1} \cos(\theta - s\xi)=
|W_t|\cos(\theta+\mbox{\rm arg}(W_t))\;,  
\end{equation} 
where $
W_t=W_t(\xi):=\sum_{s=0}^{t-1}e^{-is\xi}$.  We denote by $n$ the 
eigenvalues of the angular momentum ${\hat\mathcal N}$. Then, in 
the ${\hat\mathcal N}$
representation, the state (\ref{ubetatpsi}) reads 
(after changing variable from $\theta$ to $\theta+$arg$(W_t)$):  
\begin{equation} 
\label{ubetatpsi1} 
\langle n \ket {\Ub^t {\psi _{\beta}}} = 
 e^{in\mbox{\rm\small arg}(W_t)}
\int_0^{2\pi}\frac{d\theta}{\sqrt{2\pi}}e^{-in\theta-ik |W_t|\cos(\theta)}   
\psi_{\beta}(\theta-t\xi-\mbox{\rm arg}(W_t))\;. 
\end{equation} 
If the initial state of the particle is a plane wave (\ref{plw}) 
of momentum $p_0=n_0+\beta_0$, 
then $\xi$ takes the constant value 
$\xi_0 = \pi \ell (2\beta_0-1)$. Substituting in (\ref{ubetatpsi1}), 
and computing the integral by means of formula (\ref{bes1}),  
the momentum distribution for the $\beta_0-$rotor   
at time $t$ is: 
\begin{equation} 
\label{momdis1} 
P(n,t|n_0,\beta_0)\;=\;J^2_{n-n_0} (k|W_t|), 
\end{equation} 
where $J_n(.)$ is the Bessel function of first kind and order $n$. 
 Using the Bessel function identity  
(\ref{bess1}),  
one computes  the expectation value of $p^2$ (or the  
energy  by dividing by~2): 
\begin{equation} 
\label{energy1} 
{\overline{p^2}}(n_0, \beta_0 ,t )= \sum_n (n+\beta_0 )^2 P(n,t|n_0,\beta_0
) =  (n_0 + \beta_0)^2 +\frac{1}{2}k^2|W_t|^2. 
\end{equation} 
Explicit computation of (\ref{effd}) yields:
\begin{equation}
\label{wut}
|W_t|\;=\;\left |\frac{\sin(t\xi_0/2)}{\sin(\xi_0/2)}\right |\;\;
\mbox{\rm if}\;\;\xi_0\neq 0\;.
\end{equation}
If $\xi_0=0$ then $|W_t|=t$.  
Then   
the distribution (\ref{momdis1}) spreads linearly in time, and the average 
kinetic energy  
increases like $k^2t^2/4$. The corresponding spread over the discrete momentum 
ladder   
is the same as for a free particle under the time evolution generated  by the 
discrete Laplacian operator. 
$\xi_0=0$ corresponds to $\beta_0=1/2+n/\ell$ mod$(1)$, $n=0,1,..,\ell-1$, 
the resonant values of $\beta$ which were mentioned 
in the beginning of the present Section. 
For any other value of $\beta$, the distribution changes in time 
in a quasi-periodic manner. It  
oscillates in time  with the approximate period $\pi\xi_0^{-1}$, inverse 
to the detuning of $\beta_0$ from the nearest resonant value. 
At any time $t$, it is negligible at $|n-n_0|>k|\csc(\xi_0/2)|$.

\subsection{Incoherent ensemble of atoms.} 
\label{incoensemble}

If the initial state of the particle is a wave packet, then it is a coherent  
superposition of continuously many plane waves with different quasi-momenta,  
which are non-resonant except for a finite set of values.  
It can then 
be proven that the asymptotic growth of energy in time is proportional to  
$k^2 t/4$ \cite{FGR02}. 
Here we consider the case when the initial atomic ensemble  
is an incoherent mixture of plane waves.  
Numerical simulations based on such choices of an initial state have  
shown satisfactory agreement with experimental data \cite{DVBCL00,DGOSBG01}. 
The initial momentum distribution shall be described by a density $f(p)$. 
We can equivalently consider an ensemble of $\beta-$rotors with 
$\beta$ distributed in $[0,1]$ with the density: 
\begin{equation} 
\label{betadis} 
f_0(\beta) = \sum_{n=-\infty}^{+\infty}f(n+\beta).  
\end{equation}
In the case when $f(p)$ is Gaussian with standard deviation 
$\sigma$ (a reasonable assumption for the Oxford 
experiments according to \cite{OSMKR00,DGOSBG01}), $f_0(\beta)$ is a 
Theta-function, and Poisson's summation formula yields: 
\begin{equation}
\label{thetafunc}
f_0(\beta)\;=\;1+2 e^{-2\pi^2\sigma^2}\cos(2\pi\beta)
+ O(e^{-8\pi^2\sigma^2})\;.
\end{equation} 
For $\sigma>1$, that is relevant for the present day experiments, 
it is practically indistinguishable from the uniform 
distribution $f_0(\beta)=1$.
Each $\beta-$rotor is described by a statistical state, 
which attaches the probability $f_0(\beta)^{-1}f(n+\beta)$  
to the momentum eigenstate $\ket {n}$.  
The momentum distribution $P(p,t)$ of the particle at time $t$ is obtained  
as follows. For any given $\beta_0\in[0,1)$,  averaging 
(\ref{momdis1}) over the different $n_0$   
of the initial distribution yields the  momentum distribution  
${\bar P}(n,t|\beta_0)$ for   
the $\beta_0-$rotor. Weighted by $f_0(\beta_0)$, this is the same as  
the momentum  
distribution $P(p,t)$ of the particle over the ladder $p=n+\beta_0$
($\beta_0$ fixed, $n$ variable).  
The on-ladder distributions  corresponding to different $\beta_0$ 
combine like a jig-saw puzzle 
in building the global momentum distribution for the particle.   
The result is a complicated function of $p$ 
that  oscillates on the scale $1/t$ (see the $\xi_0-$ dependence of  
(\ref{momdis1}),(\ref{wut})). Nevertheless, in the average, this distribution 
evolves   into a steady state distribution. 
This may be shown either by time-averaging, or by coarse-graining, as done
in the following subsections.

\subsubsection{Time-averaged distribution.} 
\label{timeaverdis} 
 
\begin{figure} 
\centerline{\epsfig{figure=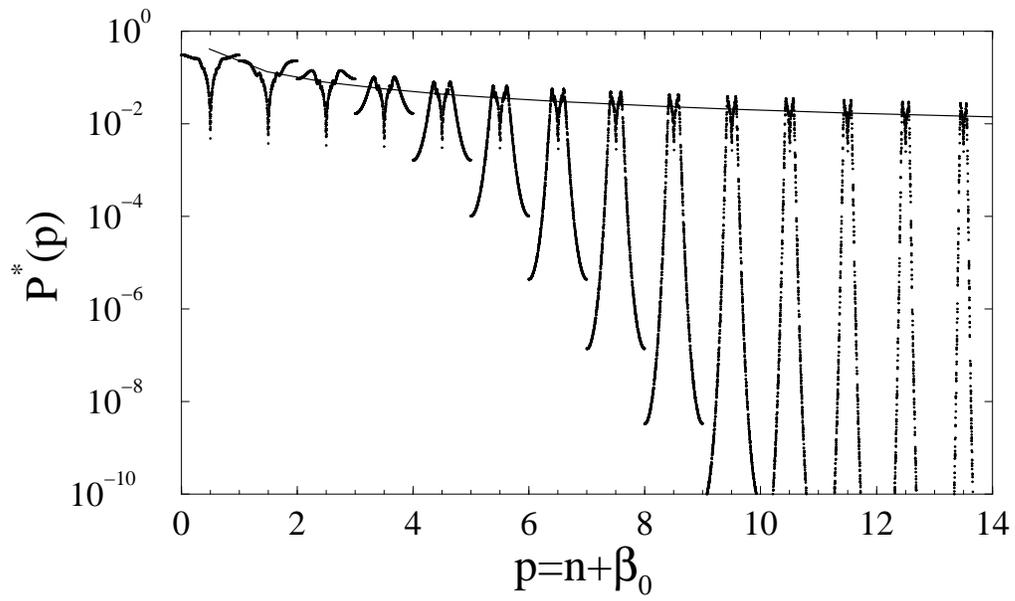,width=8cm,angle=270}} 
\caption{Time averaged stationary 
momentum distribution $P^*(p)$ for a uniform  distribution 
of initial momentum  $p_0=\beta_0$ in  $[0,1)$,  
obtained by plotting~(\ref{pstat1}) 
vs. $p=n+\beta _0$. The solid line drawn through the peak tops   
corresponds to a decay $\propto1/n$.
} 
\label{fig1} 
\end{figure}

As $t\to \infty$, the time average of the distribution (\ref{momdis1})  
tends to $0$ whenever $\beta_0$ has a resonant value, because the ballistic 
flight of atoms with a resonant quasi-momentum 
results in vanishing probability of being found  
in any {\it finite} momentum interval. For any non-resonant value of $\beta_0$, 
the distribution (\ref{momdis1}) tends 
in time-average to a distribution $P^*(n|n_0,
\beta_0)$. In fact, for non-resonant $\beta_0$ the time average of   
(\ref{momdis1}) is just the average of $J^2_{n-n_0}(k\sin(\alpha) 
\csc(\xi/2))$ along the trajectory   
of the shift $\alpha \to \alpha + \xi_0/2 ($mod$2\pi)$, 
starting from $\alpha =0$.  If $\beta_0$ is a non-resonant 
rational,  
then the trajectory is periodic and the time average trivially converges.   
If $\beta_0$ is irrational, the shift is ergodic, so    
the time average converges to the phase average. 
The time-averaged momentum distribution for the $\beta_0$-rotor 
is thus given, for irrational 
$\beta_0$, by the following integral expression: 
\begin{equation} 
\label{pstat1} 
P^*(n|n_0,\beta_0)= \frac{1}{2\pi} \int_0^{2\pi}d\alpha\; J^2_{n-n_0}  
(k \sin (\alpha) \csc(\pi\ell(\beta_0 -\frac{1}{2}))). 
\end{equation} 
It follows that the momentum distribution $P(p,t)$ for the 
atom converges, in time average, to a steady-state 
distribution $P^*(p)$. This is obtained by averaging over $n_0$ the 
time-averaged distribution $P^*(n|n_0,\beta_0)$ 
discussed above for the $\beta-$rotors, and by combining the on-ladder 
distributions 
$P^*(n|\beta_0)$ thus obtained.   
Note that the {\bf rhs} 
of (\ref{pstat1}), albeit a continuous function 
of $\beta_0$, does not necessarily coincide   
with the lhs when $\beta_0$ is rational. Hence $P^*(p)$ has, 
strictly speaking, a dense set of  
discontinuities 
 formed by values of  $p$ with a  rational non-resonant fractional part 
$\beta$. Nevertheless 
(\ref{pstat1}) is adequate   for computing   
averages of smooth functions of $p$. 
A time-averaged distribution (\ref{pstat1}) is shown in Fig.~\ref{fig1} 
for $\ell=1$ and for 
the case of an initially flat momentum distribution in $[0,1)$. 
In this case $n_0=0$, and the integral (\ref{pstat1}) was numerically 
computed for each $n,\beta_0$.
The distribution is not seen to vanish at $n+1/2$ 
because the computational grid in $p$ was chosen at random, 
thus avoiding  simple rational numbers. 
Discontinuities at integer $p$ are clearly visible. They are due 
to the discontinuity  of the initial momentum (box) distribution at 
$p=0,1$,  
and are not related to the above discussed ones.    
In the  momentum intervals  
$n<p<n+1$ with $n \gg k$ the  distribution 
$P^*(p)$ is negligibly small  
except in a neighbourhood of width $2/n$ around $n+1/2$. This is due to the 
faster-than-exponential decay of the  
Bessel functions when their  order increases beyond the argument.
At $p=n+1/2$ the distribution is exactly zero. 
Thus in all intervals $n<p<n+1$, $P^*(p)$ has the form of two narrow peaks, 
situated symmetrically wrt $n+1/2$. As shown in the next subsection, 
the total probability in such intervals decays like $1/n^2$, so the 
height of the twin peaks decays like $1/n$, as is indeed clear from
Fig.~\ref{fig1}.   
Cases with $\ell>1$ may be discussed along similar though not identical 
lines. \\

\subsubsection{Coarse-grained distribution.} 
\label{coarsedis} 
 
The second way of removing the fast oscillations is replacing $P(p,t)$ 
in each interval $n<p<n+1$ by its integral $P_n(t)$ over  
that interval. This corresponds to using a bin size $2\hbar k_L$ for  
the observed distributions.   
Assuming  $f(p)$ to be coarse-grained itself, 
the new distribution is approximately computed in the form:  
\begin{equation} 
\label{cgdis} 
P_n(t)=\sum\limits_m M_{n-m}(t)f(m), 
\end{equation} 
where 
\begin{eqnarray}
M_n(t)&=&\int _0^1 d\beta\;J^2_n\left( k|W_t|  
\right)=\int_{-\pi}^{\pi}\frac{dx}{2\pi}J^2_n(k\sin(tx)\csc(x)) 
\nonumber\\ 
&=&
\int_{-\pi}^{\pi}\frac{dx}{4\pi^2}
\sum\limits_{r=0}^{t-1}\frac{2\pi}{t}J^2_n
\left( k\sin(x)\csc(xt^{-1}+2\pi rt^{-1}) \right). 
\label{fourcg} 
\end{eqnarray} 
In the limit when $t\to\infty$ and  $2\pi r/t\to\alpha$,  
the sum over $r$ approximates  
the integral over $\alpha$, and 
(\ref{fourcg}) converges to  
\begin{equation} 
\label{fourdis} 
M_n^*=(2\pi)^{-2}\int_{-\pi}^{\pi}dx\int_0^{2\pi} 
d\alpha\;J^2_n(k\sin(x)\csc(\alpha))\;. 
\end{equation}
The steady-state coarse-grained distribution $P^*_n$ is then obtained 
by replacing (\ref{fourdis}) in (\ref{cgdis}). Note that it is 
the coarse-grained distribution itself, and not just its 
time-average, that converges to the steady-state distribution.
We do not know if the double integral may be computed in closed form. In  
\ref{appdis} we prove the following (non-optimal) estimate, valid for $N>k$:
\begin{equation}
\label{estdist}
\sum\limits_{|n|\geq N}M_n^*\;\leq\;
2\left(\frac{ke}{16}\right)^{\frac{2N}{2N+1}}
N^{\frac{1-2N}{1+2N}}\left(2+\frac{1}{N}\right)\;.
\end{equation}
Using this estimate it  
is easy to compute  that for $k>1$ the 
total probability 
carried by states $|n|>4k$ is not larger than $0.31\ldots$; so, at large 
$k$,  the distribution is rather narrow as compared to the exponentially 
localized distribution which is observed far from resonance, because 
the width of the latter scales like  $k^2$. 
In \ref{appdis} we further prove that the   distribution (\ref{fourdis}) 
has the following large-$|n|$ asymptotics:
\begin{equation}
\label{decdis}
M_n^*\;\sim\;\frac{4k}{\pi^3 n^2}\;\;\mbox{\rm as}\;\;|n|\to\infty\; .
\end{equation}
Such a decay carries over to the coarse-grained distribution   
$P^*_n$ whenever $f(p)$ is fast decaying (e.g., like a Gaussian). 
 
\begin{figure} 
\centerline{\epsfig{figure=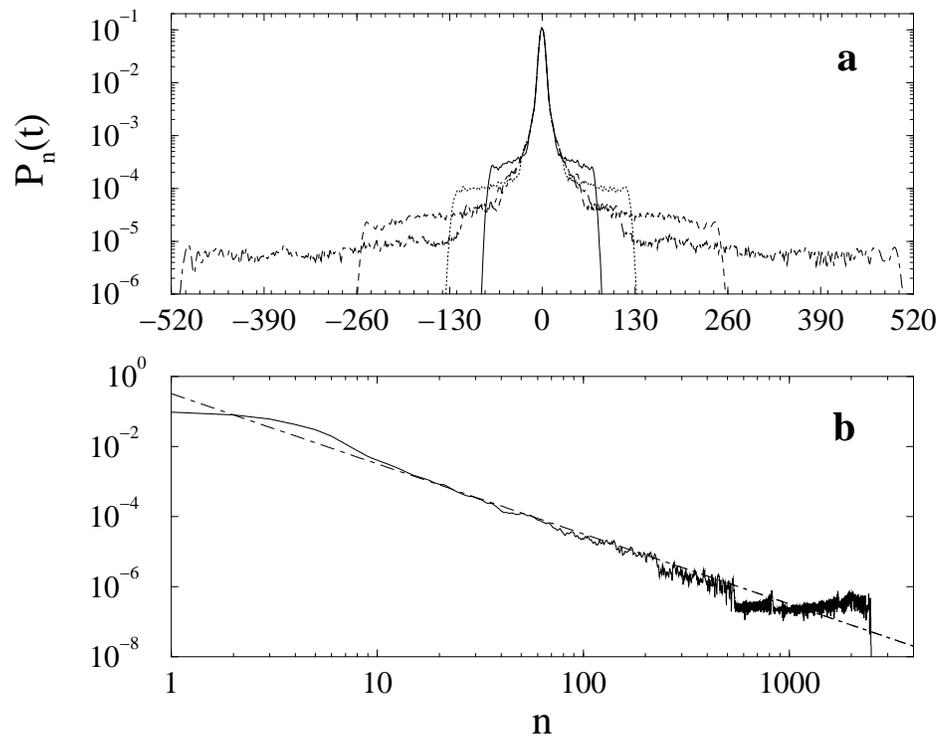,width=10cm,angle=270}} 
\caption{Evolution of coarse-grained momentum distributions~(\ref{cgdis}) 
at resonance $\tau =2\pi $, $k=0.8\pi$ for an ensemble of $10^4$ 
atoms, without decoherence. The initial momentum distribution is 
centred Gaussian, with rms deviation 
$\sigma \simeq 2.7$. 
(a): distributions for $t=30$ (solid lines), $t=50$ (dotted),  
$t=100$ (dashed), $t=200$ (dashed-dotted).  
(b): 
doubly logarithmic plot of the distribution at $t=1000$    
(solid line), compared to the asymptotic formula
$4k/(\pi^3 n^2)$ (dashed-dotted line) (\ref{decdis}).
No cut-offs are used, so the distributions 
characterize the ideal behaviour of an ensemble of $\delta$-kicked particles.
} 
\label{fig2} 
\end{figure}

The convergence of the coarse-grained distributions to the steady-state  
distribution is illustrated in Fig.~\ref{fig2}, that  
were produced by numerically simulating the evolution of  
statistical ensembles of particles, with an 
initial Gaussian  momentum distribution.  
The central part of the distribution quite early stabilizes in the  
final form of a narrow peak of width $\sim k$.   
Away from this peak, the algebraic tail  
$\propto n^{-2}$ develops over larger and larger momentum ranges  
as time increases  
in the wake of two symmetric, tiny  
``ballistic peaks''\footnote{This
denotation is borrowed from \cite{OSMKR00}, 
where experimental observation of such 
structures was reported.}, that move away linearly in time.
The fall of the distribution  is quite steep past such peaks. 
This is easily understood from the first 
equation in (\ref{fourcg}): since $|\sin(tx)\csc(x)|\leq \pi t/2$, 
at $n>\pi kt/2$ the integrand decays faster than exponentially. 
The distribution in Fig.~\ref{fig2} (b) has stabilized 
to the limit distribution over a broad momentum range.  
Apart from the far tail, where the moving peak structure 
is still apparent,  
the distribution follows  the asymptotic   decay (\ref{decdis}) 
already for $|n|\gtrsim15$. Hence, using (\ref{decdis}),  
the total probability on states $|n|>40$ is  $\sim 8\times 10^{-3}$.  
\\
From the above analysis, we obtain that all moments 
of $p$ of order $\geq 1$   
diverge as $t\to\infty$,  
in spite of the onset of the stationary distribution,   
due to the slow algebraic decay of the latter.  
For the case of the 2nd moment, which is (apart from a factor 2)  
just the mean energy of the ensemble, the growth is actually {\it linear in 
time}, as we shall presently show. 
 
\subsubsection{Average kinetic energy.} 
\label{avereng} 
 
The mean kinetic energy of an ensemble of rotors at time $t$ is obtained by 
averaging (\ref{energy1}) over the initial momentum distribution:
\begin{equation} 
\label{energyens} 
{\cal E}\{\bar E(t)\}= {\cal E}\{\bar E(0)\}
+\frac{k^2}{4} \int_0^1 d\beta f_0(\beta)  
\frac{\sin^2(t \pi\ell (\beta -1/2))}{\sin^2(\pi\ell (\beta -1/2))}. 
\end{equation} 
As $t\to\infty$, the fraction in the integrand, multiplied  by $\ell/t$, 
tends to a periodic 
$\delta$ function of $(\beta-1/2)$ with period $1/\ell$. Thus 
(\ref{energyens}) has the following $t\to\infty$ 
asymptotics:
\begin{equation} 
\label{energyensfin} 
{\cal E}\{\bar E(t)\}\sim{\cal E}\{\bar E(0)\}+ \frac{k^2t}{4\ell} 
\sum\limits_{j=0}^{\ell-1}f_0(\beta^{(j)}). 
\end{equation} 
where $\beta^{(j)}=1/2+j/\ell$ mod$(1)$ are the resonant quasi-momenta.
In the case when $f_0(\beta) \equiv 1$, i.e., 
it is uniform in $[0,1)$, this formula 
is exact for {\it all} times $t$ (by (\ref{intsinc})).  
It is practically  exact at all times  
for an initial Gaussian distribution with width  
$\sigma >1$ around the origin; for, indeed,   
$f_0(\beta)$ is nearly uniform in that case, as noted   
in section~\ref{incoensemble}. Higher order energy moments may be  
likewise computed. For large, finite $t$ the 
$n^{-2}$ decay of the distribution is truncated at $n\sim kt$ and is 
thereafter replaced by a much faster decay. Consequently, the variance 
of energy increases like $t^3$. The increase of other moments may 
also be estimated in this way.

\section{Dynamics near to Resonance.} 
\label{dynres} 
 
The experimental observation of quantum resonances is mainly based on  
measuring the average energy of the kicked atoms after a fixed time 
$t_{obs}$.  
For all sufficiently irrational values of the kicking period $\tau$,  
the theory of the ideal kicked rotor model predicts localization,  
that is, on increasing $t_{obs}$ beyond a {\it break-time} $t^*$,  
the observed energy values should not increase any more. 
In contrast, resonant values of $\tau$ lead to unbounded growth  
of energy\footnote{Unbounded growth was also proven 
for a dense set of close-to-commensurate values of $\tau$
\cite{CG84}. Extremely long times are however required to resolve 
such arithmetic subtleties.}
 with $t_{obs}$, as analyzed  in previous sections   
for the particular cases $\tau=2\pi\ell$. Hence, if $t_{obs}$ 
is significantly larger than $t^{*}$, then a scan of  
the measured energy versus the kicking period $\tau$ yields plots  
similar to the (numerically obtained) one  shown in Fig.~\ref{fig3}(a),  
where peaks are clearly observed at the resonant values $\tau=2\pi\ell$.  
For obvious continuity reasons, such peaks have a width, determined  
by the finite value of $t_{obs}$. In the ideal case, they would shrink 
on increasing $t_{obs}$, and further, 
narrower peaks associated with higher-order resonances would appear. 
In this section we derive a description of  
the structure of the peaks around $\tau=2\pi\ell$, based  
on a finite time, small-$\ep$ asymptotics, where $\tau=2\pi\ell+\ep$.  
This technique was introduced in \cite{FGR02b,FGR02}. We in particular
find that the width of the resonant peak scales like 
$(kt^2_{obs})^{-1}$, so that
at large $t_{obs}$ the peak is much narrower than the naive expectation 
$\propto 1/t_{obs}$. 
 
\subsection{$\ep-$quasi classical asymptotics near resonance.} 
\label{epsclass4} 
 
We  rescale $k=\tilde k/|\epsilon |$, 
and define  
\begin{equation} 
\label{epsscal} 
\Ih = |\epsilon | \Np = -i|\epsilon | \frac{d}{d\theta}\;,~~ 
\Hb (\Ih,t) = \frac{1}{2} \mbox{sign}(\ep) \Ih^2 + \Ih (\pi \ell +\tau \beta).
\end{equation} 
Then the  Floquet operator for the $\beta-$rotor  
may be rewritten in the form: 
\begin{equation} 
\label{onecycleeps} 
 \Ub (t)\;=\;e^{-\frac{i}{|\ep |}  
\tilde k\cos({\hat \theta})}\;e^{-\frac{i}{|\ep|}\Hb}. 
\end{equation} 
If $|\epsilon|$ is regarded as the Planck's constant, then 
(\ref{epsscal}),(\ref{onecycleeps}) is the formal quantization 
of either of the  following classical  maps: 
\begin{equation} 
\label{clmap} 
I_{t+1}=I_{t}+{\tilde k}\sin (\theta_{t+1})\;\;,\;\; 
\theta_{t+1}=\theta_t\pm I_t+\pi \ell+\tau \beta\;\;\mbox{\rm mod} 
(2\pi)\, 
\end{equation} 
where $\pm$ has to be chosen according to the sign of $\epsilon$. 
We stress that ``classical'' here is not 
related to the $\hbar\to 0$ limit but to the limit $\epsilon\to 0$ instead. 
The small$-|\epsilon|$ asymptotics of the quantum 
$\beta-$rotor is thus equivalent to a quasi-classical approximation    
based on the ``classical'' dynamics  (\ref{clmap}), that will be 
termed  {\it $\ep$-classical} in the following. 
Changing variables to $J=\pm I+\pi\ell+\tau\beta$, 
${\vartheta}=\theta+\pi(1-\mbox{\rm sgn}(\ep))/2$ turns the maps    
(\ref{clmap}) into a single Standard Map, independent of the 
value of $\beta$:
\begin{equation}
\label{epsmap}
J_{t+1}=J_t+{\tilde k}\sin (\vartheta_{t+1})\;\;,\;\;
\vartheta_{t+1}=\vartheta_t+J_t\;.
\end{equation}
This will be called the $\ep$-classical Standard Map ($\ep$SM) in what follows.
In Fig.~\ref{fig4} quantum energy curves vs. $\tau$ in a neighbourhood  
of $\tau=2\pi$ are compared with energy curves computed using 
the $\ep-$classical map 
(\ref{clmap}). 
For any given particle in the initial ensemble, the map (\ref{clmap})  
with $\beta$  
equal to the quasi-momentum of the particle was used to compute  
a set of trajectories started at  
 $I=n_0 |\ep|$ with homogeneously distributed $\theta\in[0,2\pi)$.  
The final energies $\ep^{-2} I_t^2/2$ at  $t=t_{obs}$ 
of the individual trajectories 
were averaged over $\theta, 
\beta,n_0$ with the appropriate weights. This is equivalent 
to using the \epsm~ in all cases, with  
different initial ensembles $J_0=const=\pm n_0|\ep|+\pi\ell
+\tau\beta_0$. As $\beta_0$ is varied, 
such ensembles sweep the full unit cell of the 
\epsm, so sampling different $\beta_0$'s 
amounts to probing different regions of the $\ep-$classical 
phase space.
The average energy $\langle E_t\rangle=\ep^{-2}\langle I_t^2\rangle/2$ 
is plotted vs. $\tau=2\pi+\ep$ in Fig.~\ref{fig4}, along with results of the 
corresponding quantal computations. 
The main qualitative features emerging of Fig.~\ref{fig4} are: 
(i) On a gross scale the curves 
are shaped in the form of a basin with a high, narrow spike  in the centre,  
closely flanked by a much smaller peak on either side.
(ii) Quantum and $\ep-$classical curves nicely agree at small $|\ep|$,  
in particular the structure of the spike  is the same.  
Their behaviour at large $|\ep|$ is qualitatively similar but 
quantitatively different. 

This overall qualitative  
behaviour may be explained in $\ep$-classical terms,  
and an approximate scaling law for the $t,k,\ep$ dependence 
of the mean energy close to resonance may be obtained, as shown in the 
next section. \\ 
  
\subsubsection{$\ep$-quasi-classical analysis 
of the resonant peaks.}
\label{analspike}

The  $\ep-$classical standard map is  different  
from the map obtained in the classical limit proper $\hbar\to 0$ 
of the kicked  rotor. In particular, if $2\pi\ell k>1$, 
then the  classical and the $\ep-$
classical dynamics  are at sharp  variance whenever 
${\tilde k}<1$. In the former unbounded diffusion  
occurs; in the latter the dynamics is quasi-integrable 
instead, and the $\ep-$classical trajectories  
remain trapped forever in-between  KAM curves. 
It is exactly the deep changes which occur in the $\ep-$classical 
phase space 
as $\tau$ is varied at constant $k$ that account for the 
energy vs. $\tau$ dependence at fixed time $t=t_{obs}$. 
In the following discussion 
we assume for simplicity an initially flat distribution of 
$p\in [0,1]$; then $I_0=0$, and $J_0=\pi\ell+\tau\beta_0$ 
with $\beta_0$ uniformly distributed in $[0,1)$. 
Without loss of generality 
we also consider $\ell=1$. Hence if $|\ep|\ll1$ 
then $J_0$ is uniformly distributed over one period (in action) 
$(\pi, 3\pi)$ of the \epsm. 

Since $J_t=\pm I_t+\pi+\tau\beta$, and $I_0=0$, the  
mean energy of the rotor at time $t$  is: 
$$ 
\langle E_{t,\ep}\rangle=\ep^{-2}\langle I_t^2\rangle/2=
\ep^{-2}\langle(\delta J_t)^2\rangle/2\;,\;\;
\delta J_t=J_t-J_0\;. 
$$
The exact quantum resonance  $\ep=0$ corresponds to the 
integrable limit of the \epsm, where $\delta J_t=0$. However, 
$\langle E_{t,\ep}\rangle$ is scaled by $\ep^{-2}$, so in order 
to compute it  at $\ep=0$ one has 
to compute $\delta J_t$ at first order in $\ep$. This is done 
by substituting the $0$-th $\ep$-order of the 2nd equation
into the 1st equation of (\ref{epsmap}). This leads to:
\begin{equation}
\label{Igrth}
\delta J_t\;=\;|\ep|k\sum\limits_{s=0}^{t-1}\sin(\theta_0+J_0 s)+
r(\ep,t)
\end{equation}
where 
$r(\ep,t)=o(\ep)$ as $\ep\to 0$ at any fixed $t$. 
The energy at time $t$ is found from (\ref{Igrth}) by 
taking squares,  averaging over $\theta_0$, $J_0$,  
dividing by $2|\ep|^2$, and finally letting $\ep\to 0$. 
With the help of (\ref{intsinc}), this calculation yields: 
\begin{equation}
\label{resgrth}
\langle E_{t,0}\rangle=\frac{k^2}{8\pi}\int_{\pi}^{3\pi}
dJ_0\;\frac{\sin^2(J_0t/2)}{\sin^2(J_0/2)}=\frac{k^2}{4}t\;.
\end{equation}
The small contribution of the initial quasi-momentum 
in the atom's energy was neglected. Apart from that, 
(\ref{resgrth}) is the same result  
as was found 
by the exact quantum mechanical calculation performed at $\ep=0$ and 
$\ell=1$  
in section~\ref{avereng} for the case of a uniform QM distribution, 
see~(\ref{energyens}).  
Thus the  $\ep$-quasi-classical approximation 
reproduces the quantum behaviour at exact quantum resonance.

The integral over $J_0$ in (\ref{resgrth}) collects 
contributions from all the invariant curves $J_0=$const. of the 
\epsm~ at $\ep=0$. Of these, the one at $J_0=2\pi$ leads to 
quadratic energy growth  
because it consists of (period 1) fixed points.  
This is called a classical resonance. It 
is responsible for the linear growth of energy (\ref{resgrth}), 
because the main contribution 
to the integral in  
(\ref{resgrth}) comes from  a small interval $\sim 2\pi/t$ of actions 
around $J_0=2\pi$. Note that $J_0=2\pi$ corresponds to $\beta_0=1/2$, 
the ``resonant'' value of quasi-momentum. It is hence seen that 
the $\ep-$quasi-classical approximation explains  the {\it quantum} 
resonances of the KR in terms of  the {\it classical} resonances of the 
Standard Map. 

We shall now   
estimate  $\langle E_{t,\ep}\rangle$ for $|\ep|>0$. 
 The $|\ep|>0$ dynamics 
is maximally distorted with respect to the $\ep=0$ one for $J_0$ in 
the vicinity of  $2n\pi$, that is, in the very region which is mostly 
responsible for the linear growth of energy at $\ep=0$.  
Being formed of period-1 fixed points,  
the $J_0=2n\pi$, $\ep=0$  invariant curves break  at $|\ep |>0$ 
as described by the Poincar\'e-Birkhoff 
theorem \cite{LL92}. The motion is then strongly  distorted 
inside regions of size (in action) $\delta J_{res}$ astride $J=2n\pi$. 
 Such regions are termed  the ``main resonances'' of the \epsm, and 
a well known estimate has $\delta J_{res}\approx 
4(k|\ep|)^{1/2}$ \cite{LL92}. Inside such regions, the approximation 
(\ref{Igrth}) fails quite quickly, so their    contribution 
$\langle E_{t}\rangle_{res}$  in the mean energy 
has to be differently estimated. 

In the remaining part of the $\ep-$classical 
phase space the motion mostly  follows KAM invariant curves, 
slightly deformed with respect to the $\ep=0$ ones, still with  
the same rotation angles. The contribution $\langle E_t\rangle_
{\mbox{\rm\tiny KAM}}$ of 
such invariant curves  
in the mean energy is therefore roughly similar to that 
considered in the integral (\ref{resgrth}), provided $J_0$ is therein  
meant as the rotation angle\footnote{Higher resonances 
appear  near all values of $J_0$ 
commensurate to $2\pi$. At small $|\ep|$ such higher-order 
resonances affect regions of phase space, that are negligibly small
 with respect 
to the main resonance. They are altogether ignored in the 
present discussion. Also note that  
structures that are small 
compared to the ``Planck constant'' $|\ep|$ are irrelevant for the 
purposes of the 
$\ep-$classical approximation. 
}. On such grounds, in order to roughly estimate 
$\langle E_{t,\ep}\rangle$ we remove from the integral (\ref{resgrth}) 
the contribution of the resonant action interval near $J_0=2\pi$, and we 
replace it by $\langle E_{t}\rangle_{res}$: 
\begin{equation}
\label {sup}
\langle E_{t,\ep}\rangle\sim\frac{k^2}{4}t-\Phi(t)+\langle E_t\rangle_
{res}\;,
\end{equation}
where
\begin{equation}
\label{phires}
\Phi(t)=\frac{k^2}{8\pi}\int_{-\delta J_{res}/2}^{\delta J_{res}/2}
dJ'\;\frac{\sin^2(tJ'/2)}{\sin^2(J'/2)}\;,
\end{equation}
and $J'$  is the deviation from the resonant value $2 \pi$. 
The contribution $\langle E_t\rangle_{res}$ may  be estimated  by 
means of  the well-known 
pendulum approximation \cite{LL92}. Near the \epsm~ 
resonance, the motion is described (in {\it continuous} time) 
by the following pendulum Hamiltonian in the canonical coordinates 
$J',\vartheta$:
\begin{equation}
\label{pend}
H_{res}=\frac12 (J')^2+|\ep|k\cos(\vartheta)\;.
\end{equation}
The resonance width  $\delta J_{res}$  is estimated  
by the separation (in action)  between the separatrices of the 
pendulum motion.
The period of the small pendulum oscillations is $2\pi t_{res}$ 
where $t_{res}=(k|\ep|)^{-1/2}$, so we use $t_{res}$  
as a characteristic time 
scale for the elliptic motion in the resonant zone.  
One may altogether remove $|\ep|$ from the Hamilton equations, 
by scaling momentum and time by factors $(k|\ep|)^{-1/2}=
4/\delta J_{res}$, 
$(k|\ep|)^{1/2}=1/t_{res}$ respectively. Therefore, 
\begin{equation}
\label{enspend}
\langle(\delta J_t)^2\rangle=\langle(J'_t-J'_0)^2\rangle\sim 
k|\ep| G(t\sqrt{k|\ep|})\;,
\end{equation}
for an  ensemble of orbits started inside the resonant zone,  
where $G(.)$ is a parameter-free function,  
whose explicit expression involves elliptic integrals.  
This function results of averaging over 
nonlinear  pendulum motions with a continuum of different periods, 
so it saturates  to a constant  value when 
the argument is larger than $\sim 1$.  At small values ($\ll 1$) of the 
argument it  
behaves quadratically.   
The contribution to the total energy is then obtained on 
multiplying (\ref{enspend}) by $|\ep|^{-2}\delta J_{res}/(4\pi)$, 
because only a fraction $\sim \delta J_{res}/(2\pi)$ of the initial 
ensemble is  trapped in the 
resonant zone. As a result 
\begin{equation}
\label{enspend1}
\langle E_t\rangle_{res}\sim \pi^{-1}|\ep|^{-1/2}k^{3/2}
G(t\sqrt{k|\ep|})\;.
\end{equation}
When $\delta J_{res}$ is small, $\sin^2(J'/2)$ may be replaced by $J'^2/4$ 
in the integrand in (\ref{phires}), leading to 
$$
\Phi(t)\sim \frac{k^2}{4}t\;\Phi_0(t\sqrt{k|\ep|})
$$
with 
$$
\Phi_0(x)\equiv\frac{2}{\pi}\int_0^x
ds\;\frac{\sin^2(s)}{s^2}\;.
$$
Replacing in (\ref{sup}), we obtain:
\begin{eqnarray}
\label{repla0}
R(t,k,\ep)&\equiv& \frac{\langle E_{t,\ep}\rangle}
{\langle E_{t,0}\rangle}\sim H(x)\equiv 
1-\Phi_0(x)+\frac{4}{\pi x}G(x)\;,\nonumber\\
&x&=t\sqrt{k|\ep|}
=t/t_{res}\;.
\end{eqnarray}
Hence  $R(t,k,\ep)$ depends on $t,k,\ep$ only through 
the scaling variable 
$x=t/t_{res}$.  
The width in $\ep$ of the resonant peak therefore scales like 
$(kt^2)^{-1}$. The scaling law (\ref{repla0}) is demonstrated  
by numerical data shown in Fig.~\ref{fig5}. 
The function $H(x)$ was numerically computed: in particular, 
$G(x)$ was computed by 
a standard Runge-Kutta  integration  of the 
pendulum dynamics (\ref{pend}).  The scaling function $H(x)$ 
decays proportional to $x^{-1}$ at large $x$, because so 
do  $1-\Phi_0$ and $4G(x)/(\pi x)$; the latter, due to the 
saturation of $G$. From Fig.~\ref{fig5} it is seen that $\Phi_0$ 
is quite slowly varying  at $x> 4$. The 
structures observed in that region are then due to $G(x)$, which  
describes the resonant island.   
Numerical computation shows that $G(x)$ 
saturates via a chain of oscillations 
of decreasing amplitude around the asymptotic value. 
These give rise 
to three local maxima in the graph of $x^{-1}G(x)$, 
followed by a chain of gentle oscillations 
in the tail. 
The first and most pronounced maximum lies in the small-$x$ region, 
and is not resolved by the scaling function $H(x)$,  
apparently because it is effaced by the rapid decay of 
$1-\Phi_0(x)$. The subsequent  
maximum and its symmetric partner at $\ep<0$ are instead 
resolved and  precisely  correspond  to the ``horns'' 
the right-hand one of which is marked by the arrow in Fig.~\ref{fig4}. 
The oscillations  
in the tail of $x^{-1}G(x)$ are also well  
reproduced in the tail of $H(x)$. 

The scaling law shows that at given $k,t$ the energy curve 
$\langle E_{t,\ep}\rangle$ vs. $\ep$ 
decays proportional 
to $|\ep|^{-1/2}$ past the   
 ``horns''. As $t$ increases the horns rise higher while moving  closer and 
closer to $\tau=2\pi$ , because they are located at 
a constant value of $x=t{\sqrt{k|\ep|}}$. 
Thus an overall $|\ep|^{-1/2}$ 
dependence eventually develops (cf., Figs.~\ref{fig4}, \ref{fig5}). \\
In the case when the smooth initial momentum distribution 
 includes values $n_0\neq 0$ and/or is appreciably non-uniform 
in quasi-momentum, the statistical weights of the various phase-space 
regions are different. Scaling in the single variable $t/t_{res}$ still 
holds, but the scaling functions $\Phi_0$ and $G$ may be different.

This analysis shows that the structure of the resonant 
peak is essentially  determined by the main resonant island 
 of the \epsm. It  neglects higher-harmonics  
resonances of the \epsm,  higher order islands, and especially 
the growth of the stochastic layer surrounding the main 
resonance. Such structures grow with $|\ep|$ and are expected to 
introduce deviations from 
the scaling law (\ref{repla0}).   
Hence this analysis is valid only  if  $|\ep|$ 
is not too close to the threshold for global chaos 
$|\ep|_{cr}\approx
1/k$. 
On trespassing this threshold, the {\it critical regime} of the \epsm~ 
is entered. No isolating KAM curve 
survives, so the energy curve  
rises in time for $|\ep|>|\ep|_{cr}$. Estimating the mean energy at relatively 
short times and for $|\ep|k<4.5$ is difficult, because 
of the coexistence of unbounded, non-homogeneous diffusion 
and of elliptic motion inside  residual  stable islands. The 
increase of the curve with $|\ep|$ at constant $t$ is due to the decreasing 
size of the latter 
islands, and to  the  rapid  increase of the diffusion coefficient 
(proportional to $(|\ep|-|\ep|_{cr})^\gamma$, $\gamma\simeq 3$  
at large enough $t$ \cite{MMP84,DF85}).

\subsubsection{Validity of the  $\ep-$classical approximation  near 
resonance.}

The $\ep-$quasi-classical approximation is trivially  exact at 
all times for $\ep=0$, as shown above. At nonzero 
$\ep$, it is  valid for not too large times $t$, and it 
is in the long run spoiled by quantum, non $\ep$-classical effects. 
At $|\ep|<|\ep|_{cr}$ the $\ep$-classical motion is bounded 
by KAM curves, so the main quantum mechanism leading to 
non-$\ep$-classical behaviour is tunnelling. Estimating the 
related time scales is a non-trivial problem, 
because the $2\pi$-periodicity in action 
of the $\ep$-classical phase space  may enhance tunnelling,  
and even result in delocalization, depending on the degree 
of commensuration between $2\pi$ and the ``Planck constant'' 
$|\ep|$. For instance, if $|\ep|/2\pi$ is rational, then 
the quantum motion will be ballistic asymptotically in time.  
This is just  the ordinary quantum resonance of the quantum 
kicked rotor (section~\ref{ensemble3}). In order that one such 
resonance with $|\ep|=4\pi p/q$ exists at $|\ep|$ less than some $|\ep|_0$, 
it is necessary that $q>4\pi/|\ep_0|$. It will show up  
after a time roughly estimated by $|\ep|$ times the inverse 
bandwidth. The bandwidth is estimated to decrease faster than 
exponentially 
at large $q$ \cite{IS80,Isr90}, so one may infer that the time of validity 
of the $\ep$-quasi-classical approximation is at least 
exponentially increasing with  $1/|\ep|$ as the 
exact resonance at $\ep=0$ is approached.  

At $|\ep|>|\ep|_{cr}$ the $\ep-$classical motion 
is unbounded, and the difference between $\ep$-classical and 
quantal energy curves vs. $\tau$ is basically set by various 
quantum localization effects, including localization by Cantori  
close to the $|\ep|_{cr}$ \cite{GFP84,FGP87}. 
As a consequence, 
if $t$ is large enough, then the 
$\ep$-classical curve lies  much higher than the quantum 
one. Nevertheless the latter still rises with $|\ep|$ at 
constant $t$, because of the growth of the localization 
length (that goes along with the 
growth of the $\ep$-classical diffusion \cite{CS8nn}).

\section{Decoherence induced  by spontaneous emission.} 
\label{spontanem5} 
Loosely speaking, decoherence induced by external noise is  
expected to drive a quantum system towards classical behaviour. 
One in particular expects the   
quantum resonances for $\delta-$kicked particles 
to be strongly impaired, whenever the noise spoils 
the conservation of quasi-momentum, because   
that purely quantal conservation law plays a key role  
in quantum resonances. This is exactly the case  when   
decoherence is induced by Spontaneous Emission (SE) 
\cite{KOSR98,DGOCS01,DGOSBG01,AGSC98,DVBCL00,WSDGTPLC02}. A beam of nearly 
resonant light of wave vector $\vec{k}_{T}$ 
may induce an internal transition in the atom. The absorbed 
photon is thereafter  
spontaneously re-emitted  in a random direction, and the 
whole process results in a random change of the 
momentum of the centre-of-mass 
atomic motion. The statistical average of this change 
equals $\hbar \vec{k}_T$.  
Although SEs are already produced at a quite small rate 
by the standing wave which produces the kicks, they have also been 
experimentally introduced in a controlled way \cite{DGOCS01,DGOSBG01} 
by switching on near-resonant, low-intensity laser beams   
immediately after each kick. The additional laser beams    
induced an intensity-dependent mean number 
$\nSE$ of SE events per atom and per kicking period between  
$0$ and  $~0.2$. During each kicking period $\tau$, the SE-inducing 
laser was active over a (physical) time $\tSE\approx 0.067\tau$. 
In our analysis we neglect the delay between 
absorptions and subsequent SEs. 
We hence assume that the atom undergoes random momentum changes 
due to SE at  a discrete sequence of random times. 
Here we are only interested in the 
projection of such  momentum changes   
onto the kicking direction\footnote{The effect of randomly induced  
deviations from the straight line is, however, an interesting question that 
should be studied separately.}. 
Both the SE times and the corresponding 
momentum changes are modelled by  
{\it classical} random variables, independent of the centre-of-mass motion 
of the atom. This assumption is reasonable as long as $\nSE$ is relatively 
small; otherwise, the atoms may, in the average, be slowed down or 
cooled - a velocity-dependent effect, which cannot be accounted for 
under this assumption. On the other hand, at the high field intensity 
of a quasi-resonant laser beam 
which is required for large $\nSE$ stimulated emission would prevail. 
  The statistical   
atom dynamics may  be studied by investigating the 
stochastic Hilbert-space evolution of the atomic 
state vector \cite{GM96,AGSC98}\footnote{
This approach was first implemented  for the study of randomized \dk~dynamics 
in ref.\cite{Gua84}.}.   
More specifically, we assume an initial incoherent mixture of plane waves  
with a distribution $f(p_0)$. We further  
assume that all the variables describing SE events are given and fixed. 
Their specification  defines a {\it realization} of the random SE events. 
Then we compute the deterministic  
evolution of the atomic state vectors up to time $t$.  
The quantum probability distribution of an observable in the final state  
depends on the chosen realization and on the initial state as well;  
averaging over both yields the final statistical distribution for that  
observable. The time  
evolution of distributions and related averages is the object of  
the following  analysis. 
 
\subsection{Kicked dynamics in the presence of SE.} 
\label{modsse} 
 
Let $|\psi(t)\rangle$  be the state vector  of the atom immediately after 
the $t-$th  kick. Let the integer $\nu_t$ denote the number of SEs  
during the subsequent SE-inducing window.   
Such events are assumed instantaneous. 
Denote $s_{j-1}$ the delay 
(in physical time) of the $j-$th event with respect to the $(j-1)-$th 
one, and $d_j$ 
the momentum change of the atom induced by the $j-$th SE.  
For notational convenience a $0-$th fictitious SE  is assumed to 
occur immediately after the $t-$th kick, and a $(\nu_t+1)$-th 
one immediately before the $(t+1)$-th kick, with $d_0=d_{\nu_t+1}=0$. 
\\
The state vector immediately after the $(t+1)-$th 
kick is, apart from an inessential phase factor, 
\begin{equation}
\label{asse}
|\psi(t+1)\rangle= {\hat S}(\delta_t)
e^{-ik\cos({\hat X})}
e^{-i\tau ({\hat P}+\chi_t)^2/2} |\psi(t)\rangle\;, 
\end{equation}
where:
\begin{equation}
\label{defisse}
\chi_t= \sum\limits_{j=1}^{\nu_t}\frac{s_j}{\tau}\sum\limits_{k=0}
^{j}d_k\;\;,\;\;\delta_t=\sum\limits_{j=0}^{\nu_t}d_j\;\;,
\;\;{\hat S}(.)=e^{i(.){\hat X}}\;.
\end{equation}
Note that the 1st and the 2nd operator on the rhs 
of (\ref{asse}) commute. The conservation of  quasi-momentum 
is broken by the  operator ${\hat S}(\delta_t)$; however, it 
may be restored by means of the substitution:
\begin{equation}
\label{tempg}
|\psi(t)\rangle={\hat S}(\delta_0+\delta_1+...+\delta_{t-1})|
{\tilde\psi}(t)\rangle
\end{equation}
which is a time-dependent, momentum shifting gauge transform; 
the resulting gauge will be termed 
the {\it stochastic gauge} in the following.
 Replacing (\ref{tempg}) in 
(\ref{asse}), and using 
$$
e^{-i\tau({\hat P}+\alpha)^2/2}{\hat S}(\gamma)=
{\hat S}(\gamma)e^{-i\tau({\hat P}+\alpha+\gamma)^2/2}\;,
$$
one easily finds:
\begin{equation}
\label{psitild}
|{\tilde\psi}(t+1)\rangle=e^{-ik\cos({\hat X})}
e^{-i\tau({\hat P}+\chi_t+\sum_0^{t-1}\delta_s)^2/2}|{\tilde\psi}(t)\rangle\;.
\end{equation}
This evolution does preserve quasi-momentum, so reduction to 
$\beta-$rotor dynamics may be performed  as described in 
section~\ref{bloch2}, leading to 
\begin{equation}
\label{betarotse}
|{\tilde\psi}_{\beta}(t+1)\rangle=\Up_{\beta}(t)|{\tilde\psi}_{\beta}
(t)\rangle\;;\;\;\;\;\Up_{\beta}(t)=
e^{-ik\cos({\hat\theta})}e^{-i\tau({\Np}+\eta_t)^2/2}\;,
\end{equation}
where
\begin{equation}
\label{qmomse}
\eta_t=\beta+\chi_t+\sum\limits_{s=0}^{t-1}\delta_s\;.
\end{equation}
In this way the stochastic evolution (\ref{defisse}) 
has been separated in two 
parts. One of these is described in (\ref{tempg}) by the 
operator ${\hat S}$; the other is the evolution (\ref{betarotse}) 
in the stochastic gauge. The former  is just 
a translation (in momentum) by the total momentum imparted 
by SE during the considered time; this part is 
{\it totally classical}, so it is the latter part 
that encodes the coherent stochastic evolution.

\subsection{A theoretical model for randomized 
kicked rotor dynamics.}
\label{model}

In experiments, SE occurs at random times within SE-inducing time windows, 
and there is one such window immediately after each kick. We shall say that 
a ``{\it SE event}'' occurs at the integer time $t$, whenever 
{\it at least one} SE occurs in the window following the $t-$th kick. In 
this way, the number of SEs may be larger than the number of SE events. 
The probability of a SE event is 
denoted $\pSE$. We then assume:
\par\vskip 0.2cm\noindent
(S1) SE events occurring at different times are statistically independent. 
Hence, the random variables $\delta_t$ $(t=0,1,2,...)$ specifying 
the total (projected) momentum change produced by SE in the $t-$th 
kicking period (eq.~(\ref{defisse})) are independent, identically 
distributed random variables. 
\par\vskip 0.2cm\noindent
The following two assumptions  are at once the simplest and the 
strongest possible ones. They were chosen because they greatly simplify 
the otherwise still possible, yet cumbersome exact solution. Their validity 
will be discussed in section~\ref{disnumr}, where it will be demonstrated 
that the main results derived below remain unchanged under less 
stringent and more realistic assumptions.   
\par\vskip 0.2cm\noindent
(S2) The finite  duration of the SE-operating windows is negligibly small 
compared to the kicking period. Hence, $s_{\nu_t} \approx \tau$ while $s_j 
\approx 0$ for $j<\nu_t$, so
that $\chi_t\approx\delta_t$ in 
(\ref{defisse}), and $\eta_t\approx\beta_0+\sum_0^{t}\delta_t$ in 
(\ref{qmomse}). 
Different SEs may occur in the same kicking period, each separately 
contributing 
to the total momentum change $\delta_t$ recorded in that period; 
nevertheless, their separation in time is neglected.  
\par\vskip 0.2cm\noindent
(S3) The occurrence of a SE event results in total randomization 
of the quasi-momentum. Given assumption S2, this is equivalent to assuming 
that the conditional distribution of the variable $\delta_t$ mod$(1)$,
given that a SE event occurs at time $t$,  
is uniform in $[0,1]$ (in our units).  
We further assume a zero mean for the distribution of $\delta_t$. 
While 
no further specification is needed for the formal elaborations 
below, in numerical simulations we 
shall in fact use  a uniform conditional distribution in $[-1/2,1/2]$  
(in our units). 
\par\vskip 0.1cm\noindent
(S4) As in the SE-free case, we assume the initial statistical 
ensemble to be an incoherent mixture of plane waves. 

\subsubsection{Random Walk in Hilbert space, at exact resonance.}
\label{RWH}
 
Assumption (S4) will not be used in this subsection.  
The  results of this subsection may therefore be used 
under different choices of the initial ensemble.
Let SE events occur after $t_0\equiv 0$ 
at integer times $t_1=\Delta_0, 
t_2=\Delta_0+\Delta_1,\ldots, t_j=t_{j-1}+\Delta_{j-1},..$. The variables 
$\Delta_j$, $(j\geq 0)$ are integer, independent  random variables. 
Under  assumption (S1) they are 
distributed on the positive integers $n$ with probabilities: 
\begin{eqnarray} 
\label{pois} 
\rho(n)=
\pSE(1-\pSE)^{n-1}\;\mbox{\rm for}\; n>0\;\; ,\;\;
\rho(0)=0\;,
\end{eqnarray} 
where $\pSE$ is the probability that at least one SE takes place in 
one kicking period.
For all integers  $t>0$  we  define  
$N_t=$max$\{j\;:\: t_j\leq t\}$, the number of SE events 
occurring  not later than   time $t$, and $N_0=0$.  The integer random 
variables 
$N_t$, $t\geq 0$ define a Bernoulli  process\footnote{This process 
should not be confused with the continuous 
Poisson process that may be used for the 
SEs occurring within {\it one} and the same kicking period; 
see subsection~\ref{disnumr}.}.

After such preliminaries, we 
set out to study the evolution in the stochastic gauge, 
as defined by eqs.~(\ref{betarotse}). The quasi-momentum $\beta$ 
of a $\beta-$rotor is constant in time, and for each  rotor 
$\eta_0=\beta$ (SE events are allowed immediately after kicks, 
and no kick occurs at $t=0$). 
In the stochastic gauge, 
the random propagator from time $0$ to time $t$ for the 
$\beta-$rotor is given 
by the ordered product 
\begin{equation}
\label{rdprop}
{\hat{\mathcal U}}_{S,\beta}(t)=
\prod\limits_{s=0}^{t-1}\Up_{\beta}(s)\;.
\end{equation}
The subscript $S$ on the lhs refers to the stochastic gauge. 
The  one-step propagators 
$\Up_{\beta}(t)$ are defined in eq.~(\ref{betarotse}).
Similar to what was done in section~\ref{singlebeta}  
at {\it exact resonance}   
$\tau=2\pi\ell$ one may write (cf. eq.~(\ref{ubetapsi})):
\begin{eqnarray}
\label{resse}
\Up_{\beta}(t)&=& e^{-ik\cos(\hat\theta)}e^{-i\xi_t{\hat{\cal N}}}
\equiv  e^{-ik\cos(\hat\theta)}
\Rp(\xi_s)   
\end{eqnarray}
where $\xi_t=\pi\ell(2\eta_t\pm 1)$. Although $\eta_t$ is not restricted 
in the interval $[0,1]$ in eq.~(\ref{betarotse}), the resonance condition 
allows for $\xi_t$ to be taken in $[-\pi,\pi]$ in (\ref{resse}).
 Under assumption (S2), 
$\xi_t$ has a constant value 
$\xise_j$ 
 in between $t=t_j$ and $t=t_{j+1}$, such that  
\begin{equation}
\label{xij}
\xise_j\equiv\xi_0+2\pi \ell  \sum\limits_{m=0}^{j}
\delta_m\;\mbox{\rm mod}(2\pi)\;\;,\;\;-\pi\leq \xise_j<\pi\;,
\end{equation}
where $\delta_m$ is the total momentum imparted by the $m-$th SE event. 
Hence,  
$\xi_t=\xise_{N_t}$ in between the $t-$th kick and the $(t+1)-$th one. 
A realization of the SE events is assigned by 
specifying the values of all the SE random variables just defined, 
which we collectively denote by the shorthand notations  
$\delta$ for the random momentum shifts and $\Delta$ for the random 
times of SE events.
Once the final observation time $t$ and 
the realization are fixed, for notational convenience we 
re-define $\Delta_{N_{t-1}}=t-t_{N_{t-1}}$. 
Replacing (\ref{xij}) in (\ref{resse}), and then in (\ref{rdprop}),
we get 
$$
\Up_{S,\beta}(t)=\prod\limits_{s=0}^{t-1}
e^{-ik\cos({\hat\theta})}\Rp(\xise_{N_s})\;.
$$
By repeated use of 
$$
e^{-ik\cos{\hat\theta}}\Rp(\xi)\;=\;\Rp(\xi)e^{-ik\cos(\theta+\xi)}\;,
$$
eq.~(\ref{rdprop}) may be rewritten in the form:
\begin{equation}
\label{rdprop1}
{\hat{\mathcal U}}_{S,\beta}(t)=\Rp(\xise_0+\xise_{N_1}+\ldots+
\xise_{N_{t-1}})
e^{-ik F(\hat\theta,\delta,\Delta,t)}\;,
\end{equation}
where:
\begin{equation}
\label{FFF0}
F(\hat\theta,\delta,\Delta, t)=\sum\limits_{s=0}^{t-1}
\cos(\hat\theta+\sum\limits_{r=0}^{s}\xise_{N_r}).
\end{equation}
We next define $\gamma_j=\sum_{m=0}^{j-1}\Delta_m\xise_m$. Replacing   
$s$ in (\ref{FFF0}) by $s=j+l$ with $j=N_s$, and summing over $j,l$ 
separately, 
\begin{equation}
\label{FFF}
F(\hat\theta,\xise,\Delta, t)=\sum\limits_{j=0}^{N_{t-1}}
\sum\limits_{l=0}^{\Delta_j-1}
\cos(\hat\theta+\gamma_j+l\xise_j)\;.
\end{equation}
We further  define:
\begin{equation}
\label{forthapp}
z_j=\sum\limits_{r=0}^{\Delta_j-1}\exp(i\gamma_j+
ir\xise_j)\;\;;\;\;Z_m=\sum\limits_{j=0}^m z_j\;\;;\;\;
W_t=Z_{N_{t-1}}\;,
\end{equation}
so finally 
$$
F(\hat\theta,\delta,\Delta,t)=|W_t|\cos(\hat\theta+
\mbox{\rm arg}(W_t))\;.
$$
Note that $W_t$ here differs from $W_t$ in eq.~(\ref{effd}). 
Let the initial state of the atom be a plane wave  of 
momentum $p_0=n_0+\beta_0$. 
 Given a realization $(\delta,\Delta)$,  
we operate on the corresponding rotor state (\ref{plw}) with the propagator 
(\ref{rdprop1}). The (random) state of the rotor 
at time $t$ is given, in 
the momentum representation, by:
\begin{eqnarray}
\label{ubetatpsi1r1}
\langle n \ket{\Up_{S,\beta_0}(t)\psi _{\beta_0}} 
&= &
e^{i\varphi_t}
\int_0^{2\pi}\frac{d\theta}{{2\pi}}e^{-i(n-n_0)\theta-ik|W_t|\cos(\theta)}\;,
\nonumber\\
\;\;\;\;\;\;\;\;\;
\varphi_t&=&(n-n_0)
\mbox{\rm  arg}(W_t)-n\gamma_{N(t-1)}\;.
\end{eqnarray}
The distribution of momenta $n$ at time $t$ 
is  
\begin{equation} 
\label{semomdis} 
P(n,t|n_0,\beta_0,\delta,\Delta, N)=J^2_{n-n_0} (k|W_t|). 
\end{equation}
This is formally identical to (\ref{momdis1}), but now 
$W_t$ depends on the initial 
quasi-momentum $\beta_0$ and on 
the realization of the SE events as well.  It  
is a stochastic process and  the random 
state (\ref{ubetatpsi1r1}) performs 
a random walk in the rotor's Hilbert space.

\subsubsection{Statistical Averages.} 
\label{momdisse} 

Computation of  statistical averages requires averaging over the SE 
random variables $(\delta,\Delta)$ and over the initial momentum 
$p_0=n_0+\beta_0$. Under our assumptions all such 
variables are classical random variables.   
Expectations (resp., conditional 
expectations) obtained by averaging over such classical variables will 
be denoted $\En\{.\}$ (resp., $\En\{.|.\}$). For instance, 
$\En\{.|p_0\}$ stands for the average over 
the SE variables alone, given the value of $p_0$ (or equivalently 
of $n_0,\beta_0,\xi_0$).

The large$-t$  behaviour of the stochastic process $W_t$ that drives 
the stochastic rotor evolution is ruled 
 by the large $m$ behaviour 
of the process $Z_m$. The properties of the latter process are completely 
determined by the assumptions (S2),(S3). Together with eq.~(\ref{xij}) 
assumption (S3) entails that the 
$\xise_j$ are mutually independent random variables, 
uniformly distributed in $(-\pi,\pi)$ (with the possible exception 
of $\xi_0$, whose distribution is defined by the initial ensemble).
This fact has the following consequences, that are derived 
in~\ref{appa}: the complex variables $z_j$  are 
pairwise uncorrelated whenever $j+k>1$; 
 moreover  $z_j,z_k$ are independent 
whenever $|j-k|\geq 2$. 
Thus the process $Z_m$  is a random walk in the complex plane, and  
the distribution of $Z_m$ approaches an isotropic 
Gaussian distribution in the complex plane as 
$m\to\infty$. On account of the properties of the Bernoulli process, 
the process $W_t$ at large $t$ has quite similar features. 

Moments of $W_t$ may be explicitly computed at all $t$. For instance
\begin{eqnarray}
\label{varw}
\En\{|W_t|^2\;|\;\Delta\}&=&\sum_{j,k=0}^{N_{t-1}} 
\En\{z_jz_k^*\;|\;\Delta\}\nonumber\\
&=&
\sum\limits_{j=0}^{N_{t-1}}\En\{|z_j|^2|\Delta\}+
2\chi(N_{t-1})\;\Rt(\En\{z_1z_0^*|\Delta\})\;,
\end{eqnarray}
where (\ref{zmom}) was used, and $\chi(.)$ is the characteristic  function 
of the strictly positive integers. 
The variables $\Delta_j$ were defined such that 
$\sum_{j=0}^{N_{t-1}}\Delta_j=t$;  
hence, with the help of (\ref{zmom}) and (\ref{zmom1}) we find:  
\begin{eqnarray}
\label{varw1}
\En\{|W_t|^2\;|\;\Delta\}&=&
{\cal M}(\Delta_0)-\Delta_0+t+2\chi(N_{t-1})\;\Rt{\cal N}(\Delta_0)\;.
\end{eqnarray}
Now $\Delta_0=t_1$ if $t_1\leq t$ and $\Delta_0=t$ if 
$t_1> t$, with $t_1$ the time of the 1st SE event. Furthermore, 
$N_{t-1}=0$ is equivalent to $t_1\geq t$, and $t_1$ is distributed 
according to~(\ref{pois}).  
Therefore averaging over the random 
SE times $\Delta$ yields: 
\begin{eqnarray}
\label{varw2}
\En\{|W_t|^2\}&=&t\;\mbox{\rm Prob}\{t_1\leq t\}
+{\cal M}(t)\;\mbox{\rm Prob}\{t_1> t\}+C(t,\pSE)\nonumber\\
&=&
t(1-\qSEt)+\qSEt{\cal M}(t)+C(t,\pSE)
\end{eqnarray}
where $\qSE=1-\pSE$, and 
\begin{equation}
\label{varw3}
C(t,\pSE)=\sum\limits_{n=1}^{t}\rho(n)
[{\cal M}(n)-n+2\Rt{\cal N}(n)]\;.
\end{equation}
With a smooth initial QM distribution (\ref{betadis})  
it is easily seen from 
the definitions (\ref{zmom}) and (\ref{zmom1}) of ${\cal M}$ and 
${\cal N}$  that
\begin{equation}
\label{remainder}
|C(t,\pSE)|\leq 3(\delta f_0)\sum\limits_{n=1}^tn\rho(n)\;,
\end{equation}
where $(\delta f_0)$ is 
the maximum of $|f_0(\beta)-1|$ in $[0,1)$.
In the case of a uniform initial QM distribution,   
${\cal M}(t)=t$ and $C(t,\pSE)=0$, so 
\begin{equation}
\label{unifvarw}
\En\{|W_t|^2\}\;=\;t\;,
\end{equation}
like in the case without SE. 
With a non-uniform initial QM distribution, letting $\pSE\to 0$ at fixed $t$ 
causes the 1st and the 3d term on the rhs of (\ref{varw2}) to vanish.
If instead $t\to\infty$ at fixed $\pSE>0$, then the 1st and the 
2nd term on the rhs approach $t$ and zero respectively, exponentially 
fast; the 3d term remains bounded according to 
(\ref{remainder}). So the result which is obtained with a uniform QM 
distribution is asymptotically approached.

\subsubsection{Growth of energy.} 
\label{engmom} 
 
Assuming that the  initial state and the  SE realization are given, 
and denoting $\deltot=\sum_0^{t-1}\delta_s$, 
the quantum expectation of the 
energy of the atom at time $t$ may be written as:
\begin{eqnarray}
\label{engy1}
{\overline E}(t)&=&\frac12\int dp\;p^2|\langle p|\psi(t)\rangle|^2
=\frac12\int dp\;(p+\deltot)^2
|\langle p|{\tilde\psi}(t)\rangle|^2\nonumber\\
&=&\frac12\int dp\;p^2|\langle p|{\tilde\psi}(t)\rangle|^2+\frac12
\deltot^2+\deltot\int dp\;p|\langle p|{\tilde\psi}(t)
\rangle|^2\;,
\end{eqnarray}
where (\ref{tempg}) was used. 
This expression has to be averaged over the initial statistical 
ensemble and over all SE realizations. Then
\begin{equation}
\label{engy2}
\En\{\deltot^2\}=D(t-1)\;\;,
\;\;\En\{\deltot\}=0\;,
\end{equation}
where $D=\En\{\delta_t^2\}$ is the mean square momentum change per period 
due to Spontaneous 
Emission. 
For an initial plane wave (\ref{plw}) 
of momentum $p_0=n_0+\beta_0$, with the help of 
(\ref{semomdis}) one finds 
\begin{eqnarray}
\label{2prod}
\int dp\;p|\langle p|{\tilde\psi}(t)\rangle|^2&=&\sum\limits_n
\int_0^1d\beta\;(n+\beta)|\langle n|{\tilde\psi}_{\beta}(t)
\rangle|^2\nonumber\\
&=&\sum\limits_n(n+\beta_0)J^2_{n-n_0}(k|W_t|)=n_0+\beta_0.
\end{eqnarray}
where $J_n(.)=(-)^nJ_{-n}(.)$ and  $\sum_nJ^2_n(.)=1$ 
 were used. Similarly, 
\begin{eqnarray}
\label{engy3}
\int dp\;p^2|\langle p|{\tilde\psi}(t)\rangle|^2
=\sum\limits_n 
(n+\beta_0)^2J_{n-n_0}^2(k|W_t|)\;,
\end{eqnarray}
whence, using (\ref{bess1}), 
\begin{equation}
\label{engy4}
\int dp\;p^2|\langle p|{\tilde\psi}(t)\rangle|^2=
\frac12k^2|W_t|^2+(n_0+\beta_0)^2. 
\end{equation}
Replacing (\ref{2prod}) in (\ref{engy1}), 
the expectation of the last term in (\ref{engy1}) vanishes due to 
(\ref{engy2}).
The expectation of (\ref{engy4}) is found with the help 
of (\ref{varw2}). Thus finally: 
\begin{equation}
\label{engyfin}
\En\{{\overline E}(t)-{\overline E}(0)\}=
\frac14 k^2[t (1-\qSEt)+{\cal M}(t)\qSEt+C(t,\pSE)]+
\frac12 D(t-1).
\end{equation} 
This result reduces to the SE-free one 
eq.~(\ref{energyens}) for $\qSE=1$.  
 The term on the rhs which includes  $k^2$ as a factor 
is the mean energy in the stochastic gauge. With a uniform QM distribution, 
it reduces to $k^2 t/4$ (cf.~(\ref{unifvarw})). 
so it  does  not contain 
any SE-related parameters, and is  in fact 
identical to the result  obtained in the SE-free case   
(for an initially uniform QM distribution). In experiments, 
$D$ is typically small (see section~\ref{disnumr}), and the QM distribution is practically 
uniform; so the growth of the mean energy is but weakly affected by SE. 
However, 
a similar, albeit cumbersome computation of higher-order moments
would reveal sharp differences, which  
reflect totally different ways  
of spreading of the momentum distribution in the two cases. 
At large time, 
this can be analyzed in detail, as shown in the next section.

Under assumption (S3), the QM distribution of an 
atom  is immediately turned uniform by the first 
SE event. The time scale for uniformization of 
the QM distribution is  then $t_c=-1/\log(\qSE)$. 
Eq.~(\ref{engyfin}) shows that for 
$t\gg t_c$ the growth of energy is linear with
the coefficient $k^2/4+D/2$, like in the case of a uniform QM distribution.  
On the other hand,  since $C(t,\pSE)$ is bounded in time, 
for  $(\delta f_0)t_c\ll t\ll t_c$ the growth of energy 
is dominated by the term ${\cal M}(t)$, which  is the same as in the 
SE-free, non-uniform  case. 

\subsubsection{Asymptotic momentum distribution.}
\label{assym}

In this section we assume $n_0=0$; averages over 
initial distributions allowing for $n_0\neq 0$ may be  easily implemented on 
the final results. 
We denote $P(p,t)$ the momentum distribution at time 
$t$. We show that, as $t\to\infty$, $P(p,t)$ approaches a Gaussian 
distribution with mean value $0$, in the sense that, for an arbitrary 
smooth function $\phi(p)$, 
\begin{equation}
\label{wsense}
\lim\limits_{t\to\infty}\langle\phi\rangle_t\equiv
\lim\limits_{t\to\infty}{\sqrt t}\int dp\;P(p{\sqrt t},t)\phi(p)
=\int dp\;\phi(p){\cal G}_{D+k^2/2}(p)\,
\end{equation}
where ${\cal G}_{\sigma^2}(p)$ denotes the normal distribution with zero 
mean and variance $\sigma^2$. 
To this end we compute: 
\begin{eqnarray}
\label{phiavg}
\langle\phi\rangle_t&=&\En\left\{\int dp\;
|\langle p|{\psi}(t)\rangle|^2\phi(p/{\sqrt t})\right\}
\nonumber\\
&=&\En\left\{\int dp\;|\langle p|{\tilde\psi}(t)\rangle|^2
\phi((p+\deltot)/{\sqrt t})\right\}\nonumber\\
&=&\En\left\{
\sum\limits_n  J^2_n(k|W_t|)\phi((n+\beta_0+\deltot
)/{\sqrt t})\right\}\;.
\end{eqnarray}
For $t\gg 1$ one may neglect corrections of order 
$1/{\sqrt t}$ in the argument of the smooth function $\phi$, so
$$
\langle\phi\rangle_t\approx\En\left\{
\sum\limits_n J^2_n(k|W_t|)\phi((n+[\deltot]
)/{\sqrt t})\right\}\;, 
$$
where  $[.]$ denotes the integer part. Asymptotically as $t\to\infty$, 
the statistics of $W_t$ are determined by the 
{\it fractional} 
parts of sums of many $\delta_s$ (cf. eqs.~(\ref{xij}),(\ref{forthapp})). 
Such sums of a large number of independent terms have a broad distribution, 
so their integer and fractional parts tend to be independent of each 
other  as the number of terms in the sums  diverges. The squared 
Bessel functions and the function $\phi$ in the last equation may then be 
separately averaged. Denoting 
\begin{equation} 
\label{checaz}
\phi_t(p)\equiv
\En\{\phi(p+[{\overline\delta}_t]/{\sqrt t})\} 
\end{equation}
we may write 
\begin{equation}
\label{discfhi}
\langle\phi\rangle_t\approx\sum\limits_n
\En\left\{J^2_n(k|W_t|)\right\}\phi_t(n/{\sqrt t})\;.
\end{equation}
As $t\to\infty$, 
the distribution of $W_t$ approaches an isotropic Gaussian
distribution in the complex plane centred at 0. 
The variance is found from (\ref{varw2}) to be $\sim t$ at large $t$. 
Hence the distribution  
of $\rho=|W_t|$ is asymptotically at large $t$ given by 
$dF_t(\rho)=2t^{-1}\rho d\rho\;\exp(-\rho^2/t)$. Consequently, 
\begin{eqnarray}
\label{sigcl1}
\langle\phi\rangle_t &\approx& 
\int_0^{\infty}dF_t(\rho)\sum\limits_n J^2_n(k\rho)\phi_t(n/{\sqrt t})
\nonumber\\
&=&2\int_0^{\infty}dx\;x e^{-x^2}\sum\limits_n
J^2_n(k x {\sqrt t})\phi_t(n/{\sqrt t}).
\end{eqnarray}
The integral over $\rho$ is a classical expectation, 
but the sum  over $n$ is a quantum expectation instead. It may be written 
as: 
\begin{equation}
\label{qexpe}
I_t(kx) \equiv \sum\limits_n J^2_n(k x {\sqrt t)\phi_t(n/{\sqrt t})=
\langle 0|{\hat{\cal K}}^{\dagger}
\phi_t( t^{-1/2}{\hat{\cal N}}}){\hat{\cal K}}|0\rangle ,\;\;
\end{equation}
where
\begin{equation}
\label{qexpe1}
{\hat{\cal K}}=e^{-ikx t^{1/2}\cos(\hat\theta)}\;.
\end{equation}
If we regard $t^{-1/2}$ as the Planck constant, then 
$t\to\infty$ is equivalent to  a classical limit. In that limit 
$t^{-1/2}\Np$ corresponds to (angular) momentum $p$, and  
${\hat{\cal K}}$  corresponds 
to $p\to p+kx\sin(\theta)$.  Therefore, 
the ``classical'' limit ($t\to\infty$) 
for the momentum distribution 
in the state ${\hat{\cal K}}\ket{0}$ is given by the distribution of $ 
kx\sin(\theta)$, with $\theta$ uniformly distributed in $[0,2\pi]$.
Replacing the quantum expectation (\ref{qexpe})  
by the 
average over the related classical distribution yields 
\begin{equation}
\label{qexpe2}
\lim\limits_{t\to\infty}I_t(kx)=
\int_{-\pi}^{\pi} \frac{d \theta}{2 \pi} \phi_{\infty}(kx\sin\theta)
=\int_{-kx}^{kx}\frac{dp}{\pi} 
\frac{\phi_{\infty}(p)}{\sqrt{k^2x^2-p^2}}. 
\end{equation}
Substituting this in (\ref{sigcl1}), and computing the elementary 
integral over $x$ gives
\begin{equation}
\label{itn}
\lim\limits_{t\to\infty}\langle\phi\rangle_t=\int_0^{\infty}dx\;x e^{-x^2} I_{\infty}(kx)=
\frac{1}{k\sqrt{\pi}}\int dp\;\phi_{\infty}(p)e^{-p^2/k^2}\;.
\end{equation}
On the other hand, by eq.~(\ref{checaz})   
$$
\phi_{\infty}(p)=\int dp'\;\phi(p-p'){\cal G}_D(p')\;,
$$
where ${\cal G}_D$ is the limit ($t\to\infty$) normal distribution 
of $\deltot/{\sqrt t}$. 
Recalling (\ref{itn}) and the definition 
of $\langle\phi\rangle_t$ given in  (\ref{wsense}) we immediately 
obtain the result claimed there.  Hence, $P(p,t)$ 
is asymptotically equivalent to a Gaussian with zero mean 
and variance $k^2t/2+Dt$. Being just the leading term 
in the asymptotic approximation as $t\to\infty$, this misses 
those terms in the exact result (\ref{engyfin}) which are bounded in time.
The way (\ref{itn}) was derived from (\ref{discfhi}) shows 
that decoherence turns the dynamics classical by causing the effective 
Planck's constant to decrease with time.   
An exact derivation of (\ref{itn}) from (\ref{discfhi}) 
is given in \ref{appb}.

\subsubsection{Discussion of the model, and numerical results.} 
\label{disnumr} 

We shall now discuss assumptions 
(S1-3); regarding (S4) see section ~\ref{incoensemble}. \\
(S1): The experimental time window for 
SE: $\tSE =0.067 \tau= 0.424$, for $\tau=2 \pi$ 
(or $4.5\mu sec$, in the Oxford experiments) is on the one 
hand very large 
compared to the time scales given by the 
Rabi frequency of the used atomic transition and by 
the inverse SE damping rate \cite{DGOCS01,DGOSBG01}. On the other hand, 
$\tSE$ is very small compared to the kicking period.  
It is therefore reasonable to neglect 
memory effects inherent to SE processes \cite{CDG92}, so this assumption  
appears legitimate. \\  
(S2): For the case when SE is induced 
by the kicking wave itself, this assumption remains valid as long as the 
$\delta-$kick approximation is valid. For the 
case of the Oxford experiments it may be to some extent supported  
by the smallness of $\tSE/\tau \simeq 0.067$. 
It should however be mentioned that the 
main results of our analysis: (\ref{engyfin}) and the 
asymptotically Gaussian distribution still 
hold in the absence of (S2), though a considerably more involved 
analysis is required. Rather than delving into such analysis, 
we support this claim by  numerical results to be 
presented below.\\ 
(S3): Complete randomization 
of quasi-momentum after each SE-inducing cycle occurs   
when the mean number of SEs per period is large, $\nSE \gg 1$:  
then effective  averaging leads to
quasi-independent $\eta_t$ in (\ref{qmomse}). If else $\nSE$ is small, then 
complete randomization requires that 
the conditional distribution of $\delta_t$ mod$(1)$ (given that 
at least one SE occurs in the $t-$th kicking period) be uniform 
in $[0,1]$.   
On the other hand  $\delta_t$  
is the  sum of a  random number of momentum changes  
due to single SEs. 
If these are assumed independent and identically distributed, then 
each of them has to be uniformly distributed in some interval of 
integer width. 
For a single transition in a 3-dimensional atom, the probability 
distribution of momentum shifts produced by  SEs 
is not isotropic \cite{MW97}. This in particular implies that  
the distribution of single SE,  
projected momentum shifts $\delta p$ is {\it not} uniform. 
In the case when the SE-inducing beam is orthogonal to the kicking direction,
it has the parabolic form:   
\begin{eqnarray} 
\label{paraboldensity} 
{\mathcal P_0}(\delta p) 
= \left\{ 
\begin{array}{r@{\quad , \quad}l} 
C\left(\frac{9}{8}-\frac{3(\delta p)^2}{2}\right) & |\delta p| \leq k_T/2k_L  
\\ 0 & \mbox{\small otherwise}, 
\end{array} \right. 
\end{eqnarray} 
where $C$ is a normalization constant, and 
$\vec{k}_T \perp \vec{k}_L$ are the (assumed to be orthogonal) 
wave vectors of the SE-inducing light and 
 of the kicking light, respectively.
This distribution is derived for a situation  
where SE from a $\Delta m =\pm1$ atomic transition is induced  
by circularly polarized light \cite{MW97}. 
The allowed change in momentum $\delta p$  
is restricted  within the interval $[-k_T/2k_L,k_T/2k_L]$, with 
$k_T/2k_L\simeq1/2$ (resulting in $C\simeq 1$) in 
\cite{DGOCS01,DGOSBG01}. With a non-uniform distribution 
such as~(\ref{paraboldensity}) some correlation is established 
between QMs in different kicking periods. 
The mean momentum change due to absorption followed by SE is 
$\hbar \vec{k}_T$ for a single SE-inducing beam with  wave vector 
$\vec{k}_T$. Our assumption of zero mean (along the kicking direction)
is justified either when $\vec{k}_T \perp \vec{k}_L$, or  
when the experimental arrangement uses two or more 
appropriately directed beams, whereby the atoms may be excited with equal 
probability. In such cases, the distribution of the projected 
$\delta p$  is more complicated (and closer to uniformity) than 
(\ref{paraboldensity}). 
Since the experiments use a large ensemble of Ce atoms, and 
SEs involve several 
hyperfine sublevels \cite{DGOSBG01,Dar02}, the assumption of a
nearly uniform distribution of momentum changes seems, however, most
appropriate.
We shall nonetheless  use (\ref{paraboldensity}) as a term of comparison in 
``type (II) simulations'' (see below) in order to test the effects 
of deviations from uniformity. Such numerical data  demonstrate  
that our  assumption of a uniform distribution in an 
exactly integer interval of allowed momentum 
changes does not affect the results, for experimentally relevant times 
at least.\\  
For the uniform distribution of $\delta p$ in the interval $[-1/2,1/2]$,  
$\langle\delta p^2\rangle=1/12$, so  the coefficient $D$ in 
(\ref{engy2}) is $D=\nSE/12$. With the distribution (\ref{paraboldensity}) 
$\langle\delta p^2\rangle=3/40$, and  
$D=\nSE 3/40$. In the theoretical model based on assumption (S2)  
the distribution of the random times at which single SEs occur  
within {\it one} kicking period is totally irrelevant, so 
$\pSE$ and $\nSE$ enter as {\it independent} parameters. They have 
to be related to each other in order to make contact with experiments. 
A seemingly natural way assumes  a Poisson distribution for the SEs 
occurring within one operating window, at least for not too large 
$\pSE$. In that case, $\pSE=1-\exp(-\nSE)$. 
\\
We have performed numerical simulations of two types. Type (I) 
used all assumptions (S1-4); type (II) had  
assumptions (S2),(S3) replaced by more realistic ones, 
allowing e.g. for free evolution in between successive SEs 
occurring in the same kicking period, and using various 
distributions of $\delta p$, such as (\ref{paraboldensity}), 
as discussed above. 
Type (I) simulations serve as a demonstration 
of the theoretical exact results, and much more as 
a term of comparison 
with type (II) simulations. The essential agreement between the two 
types demonstrates that our theoretical  conclusions remain valid, under 
less stringent assumptions. Both types of numerical results were 
obtained by independently evolving rotors in a given Gaussian 
ensemble, and by 
incoherently averaging the final results. 
Random SE events were simulated as follows. After choosing values 
for $\tSE$ and $\pSE=1-\exp(-\nSE)$, 
random SE times were generated in each kicking period 
from a Poisson distribution with the characteristic 
time $\tSE/\nSE$ within the time window $(t\tau, t\tau+\tSE)$.  
To each 
random  time a random momentum jump was associated, from the chosen 
distribution (uniform or parabolic). In type (I) simulations, such jumps were 
added to the quasi-momentum the rotor had at (integer) time $t$. 
The integer part of the result determined a corresponding shift in the 
computational basis of angular momentum eigenstates. 
The fractional part 
was used as quasi-momentum for a full one-period free rotor evolution. 
In type (II), free evolution was allowed in between subsequent SE times.  
In all cases {\it the computational basis of momentum eigenstates 
was chosen as large as possible} in order to model as faithfully 
as possible the ideal models analyzed in previous sections. \\
Fig.~\ref{fig6} shows a long-time 
plot for different rates $\pSE=0.05\ldots 0.2$, and for the two cases: 
type (I) with SEs happening immediately after the 
kicks (a), and type (II) with SEs within a finite time 
window ($\tSE=0.067\tau=0.424$ \cite{DGOSBG01}) (b). For $\pSE=0.2$ data 
is given in Fig.~\ref{fig6} (c) 
also for the parabolic distribution~(\ref{paraboldensity}).  
The energy growth is {\it in all cases} linear with 
the predicted slope $D_{dec}\simeq k^2/4 + D/2$,  
as discussed above (see eq.~(\ref{engyfin})). 
Fig.~\ref{fig7} presents the 
coarse-grained momentum distributions $P_n(t)$ (see section~\ref{coarsedis})
defined as the probability that the momentum $p$ 
of an atom at time $t$ lies in $[n,n+1)$ (in our units).
They are computed  
for $\tau =2\pi$ and different SE rates, SE events immediately after 
the kicks in type (I), within the time window $\tSE=0.424$ with a uniform 
distribution, or SEs immediately after kicks with distribution 
(\ref{paraboldensity}) in type (II). With added decoherence, the distribution 
keeps spreading  as a whole all the time, looking more and more Gaussian-like 
while it flattens out. Further remarks are given in section~\ref{exp6}.
Apart from statistically induced fluctuations in the wings of the distributions
no significant difference between the different simulations
(Figs.~\ref{fig6}-\ref{fig7}) is detectable, and our conclusions 
from the preceding subsections remain valid in all cases.  
We conclude that the results obtained in the work are not very sensitive to 
assumptions (S1-4) that made the analytical treatment possible.  

\subsection{$\ep-$ quasi-classical approximation, in the presence 
of SE.}
\label{epswithnoise}

The $\ep-$quasi-classical approximation introduced in 
section~\ref{dynres} for the 
study of the coherent nearly resonant quantum motion is easily 
adapted to  the model in the presence of  SE, because the 
effects of SE were  modelled by a totally classical noise. 
In the stochastic gauge, 
the $\ep-$classical approximation may be implemented 
in the $\beta-$rotor propagators (\ref{betarotse}) much in the same way as 
in subsection~\ref{epsclass4}. The resulting $\ep-$classical 
map corresponding to (\ref{clmap}) is:
\begin{equation}
\label{clmapse1}
I_{t+1}=I_t+{\tilde k}\sin(\theta_{t+1})\;\;,\;\;
\theta_{t+1}=\theta_{t}\pm I_t+\pi\ell+\tau\eta_t\;.
\end{equation}
We now exploit assumption (S2) and write $\eta_t=\beta+\deltot$, 
where $\deltot=\sum_{s=1}^t\delta_s$ is the total momentum imparted by SE up 
to 
time $t$. In order to turn off the stochastic gauge, we need to recover 
the accumulated SE momentum change, hence we change  
variables to $I^*_t=I_t+|\ep|\deltot$. The momentum of the atom at time $t$ 
is then $|\ep|^{-1}I_t^*+\beta$, 
where $\beta$ is the initial quasi-momentum. Denoting $\eta^*_t
=\eta_t+\ep\beta/(2\pi\ell)$, 
a straightforward computation yields:
\begin{eqnarray}
\label{clmapse}
I_{t+1}^*&=&I_t^*+|\ep|\delta_{t+1}+{\tilde k}\sin(\theta_{t+1})\;,\nonumber\\
\theta_{t+1}&=&\theta_t\pm I_t^*+\pi\ell+2\pi \ell  \eta_t^*\;, 
\nonumber\\
\eta_{t+1}^*&=&\eta_t^*+\delta_{t+1}\;,\nonumber\\
\eta_0^*&=&\tau\beta/(2\pi\ell)\;.
\end{eqnarray}
The $\delta_t$ are independent random variables, whose distribution  
is determined by the statistics of Spontaneous Emission. 
Numerical simulations 
of such noisy $\ep-$classical maps are shown in Fig.~\ref{fig8}, and 
very well match with the quantum computations at small $|\ep|$. 
The structure of the resonant peak in the presence of SE may be 
analytically  analyzed, using ideas developed in section~\ref{dynres}. 
Under the substitution $J_t=\pm I^*_t
+\pi\ell+2\pi \ell \eta_t^*$ 
the map (\ref{clmapse}) reduces to a noisy   \epsm~, which 
differs from the \epsm~ by a random shift $\tau\delta_t$ of the action 
$J$ at each step. 
 We assume an initially uniform QM distribution.
At any SE time 
$t_j$, the distribution of the ensemble in the phase space of 
the \epsm~ is reshuffled by the random action change. 
Under the assumption of homogeneous distribution 
of single SEs  in an interval of integer length, the resulting 
distribution of  $J$mod$(2\pi)$ is approximately 
homogeneous over the unit cell 
of the \epsm. Such randomization may be assumed to wash out correlations 
between the past and the subsequent random dynamics. 
Hence the scaling (\ref{repla0}) may be used to write the 
energy at time $t$ as 
$$
\langle E_{t,\ep}\rangle\sim\frac{k^2}{4}\langle \sum\limits_{j=0}^{N_{t-1}}
\Delta_j\;H(\Delta_j/t_{res})\rangle+\frac12 D'\nSE t
\;,
$$
where $\langle.\rangle$ stands for 
average over all the Bernoulli realizations of the times of SE events, 
 $\nSE$ 
is the average number of Spontaneous Emissions per period, and 
$D'=\nSE^{-1}\langle\delta_t^2\rangle$ is the mean square momentum 
imparted by a single SE. 
If  $t_c$ is  sufficiently large compared to $1$,  one 
may  replace the Bernoulli  process by the  continuous time 
 Poisson process 
with the characteristic time $t_c=-1/(\log(1-\pSE))=1/\nSE$. 
This process  has the delays $\Delta$ distributed with the density 
$t_c^{-1}\exp(-\Delta/t_c)$. Its statistics reduces to that 
of the unit Poisson process ($t_c=1$) by just rescaling all times by the 
factor $1/t_c$. This entails 
\begin{eqnarray}
\label{pois12}
\langle \sum\limits_{j=0}^{N_{t-1}}
\Delta_j\;&H&(\Delta_j
/t_{res})\rangle\approx 4t_c Q(t/t_c,t_c/t_{res})\;,
\end{eqnarray}
where 
\begin{equation}
\label{scalQ}
Q(u,v)\equiv\frac14\langle\sum\limits_{j=0}^{N_{u}^1}
\Delta_j^1\;H(\Delta_j^1 v)
\rangle\;.
\end{equation}
The superscript $1$ specifies that the average is now over the realizations 
of the unit Poisson process:
each realization has 
the continuous time interval $[0,u]$ divided in subintervals 
$\Delta_j^1$ by a random number 
$N_u^1$ of Poisson events. 
We are hence led to the following scaling law:
\begin{equation}
\label{replace}
\langle E_{t,\ep}\rangle\sim D'\frac{t}{2t_c}+k^2 t_c Q(\frac{t}{t_c},
\frac{t_c}{t_{res}})
\end{equation} 
or, equivalently,
\begin{equation}
\label{scalse}
\frac{2\langle E_{t,\ep}\rangle-D't/t_c}{2k^2t_c}\sim Q(u,v)\;\;,\;
u=t/t_c\;,\;v=t_c/t_{res}\;.
\end{equation}
The scaling function $Q(u,v)$ may be explicitely written in terms of the 
function $H(x)$ by a routinely calculation reported in~\ref{poisson}:
\begin{equation}
\label{scalQ1}
4Q(u,v)=uH(uv)e^{-u}+
\int_0^udx\;e^{-x}xH(xv)(2+u-x)\;.
\end{equation} 
Limiting behaviours of the scaling function 
$Q(u,v)$ immediately follow from this equation, or 
from (\ref{scalQ}) itself. On one hand, for $u=t/t_c\gg 1$ 
the rhs in (\ref{scalQ}) is a sum of 
a large number $\sim t/t_c$ of terms.  In that limit, 
such terms are quite weakly correlated and may be independently averaged, 
leading to:  
\begin{equation}
\label{lim1}
u\gg 1\;:\;\;\;Q(u,v)\sim \frac14u\int_0^{\infty}dx\;H(vx)x
e^{-x}\;.
\end{equation}
On the other hand, for $t/t_c\ll 1$, 
 the sum reduces to the single term $j=0$, with $\Delta_0^1=t/t_c$; hence
\begin{equation}
\label{lim2}
u\ll 1\;:\;\;\;Q(u,v)\sim \frac14u H(uv)\; .
\end{equation}
In particular, (\ref{lim2}) shows that 
(\ref{replace}) coincides with (\ref{repla0}) in the
SE-free limit $t_c\to\infty$. In the opposite limit, (\ref{lim1}) 
shows that if $k$ is fixed then the width in $\ep$ of the resonant 
spike does not shrink any more  with time when $t\gg t_c$, and its 
width thereafter scales like $(t_c^2k)^{-1}$. The spike 
 is therefore erased (that is, it is absorbed in the background) 
in the strong noise limit $t_c\sim 1$. \\
If $f_0(\beta_0)$ is smooth but not uniform, 
the scaling in absence of SE of the form~(\ref{repla0}) holds but 
with a different 
scaling function $H$ as explained in subsection~\ref{analspike}.   
Therefore the arguments of the present subsection leading to 
(\ref{replace}) should hold also in this case.

Numerical simulations in Fig.~\ref{fig9} 
satisfactorily support this scaling law. 
Data are obtained in a similar manner as for the case without SE; however, 
one  of the parameters $u,v$ is varied, while keeping fixed 
either the other parameter or the ratio $u/v$. The theoretical 
scaling function $Q(u,v)$ was numerically computed using in (\ref{scalQ1}) 
the function $H(x)$ computed as described in section~\ref{analspike}. 

\section{Reconciliation with experimental results.} 
\label{exp6} 
 
In the presence of SE induced decoherence, 
experimentally measured energies  at fixed observation time 
$t_{obs}$ were found to 
exhibit resonant peaks near the resonant values  $\tau=2\pi\ell$ (integer
multiplies of the half-Talbot time \cite{DGOCS01,DGOSBG01,DB1997,Goo1996}) 
that were {\it higher} than in the SE-free case. Numerical support for this 
observation was recently given in \cite{DPLT02}. Such observations may 
have been suggestive of an 
enhancement of quantum resonances due to decoherence: however, such a  
phenomenon has no match in the theory 
developed in the previous sections. This paradox will be resolved 
in this section. It will be shown that 
certain restrictions, that are unavoidably
present in real experiments, depress the ideal 
resonant behaviour in a way, that is most  severe {\it in the absence   
of SE}. So the explanation rather lies with the experimentally measured,  
SE-free peaks being {\it lower} with respect to the ideal case, 
than with the SE ones being higher. 

The most important  experimental features not taken into account 
in the foregoing  theoretical analysis are: 
\par\vskip 0.2cm\noindent
(I) experimental kicks are not $\delta-$like.  
The ideal  model is then only valid as long as  
the distance 
travelled by the atoms over the finite duration $t_p$ of the kick 
is much smaller than 
the spatial period of the potential ($t_{p} = 0.5 \mu sec$ in 
\cite{DGOCS01,DGOSBG01}). In our units
$\tau_p=t_p\hbar(2k_{L})^2/M\simeq 0.047$. 
At large momenta this requirement is violated, and the atomic motion  starts  
averaging  over the potential. The small momentum regime is practically 
not affected by the replacement of the $\delta-$function by a pulse of finite 
width. A proper theoretical description demands 
the $\delta-$kicks in the atomic Hamiltonian to be replaced by 
appropriately shaped pulses \cite{BFS86}. For  large momenta the pulses act
 adiabatically 
(if they are smooth), leading to trivial classical and 
quantum localization \cite{BFS86,KOSR99,DVBCL00}. The classical phase space  
is then filled  by KAM tori beyond some large momentum 
value $n_{ref}$. For smooth pulses, 
the momentum $n_{ref}$ is inversely 
proportional to the duration of the pulse $\tau_p$, and the 
pre-factor of the proportionality depends on the shape of the pulse
\cite{BFS86}. 
The translational invariance in momentum required for quantum 
resonances is thus destroyed. The atom dynamics  
mimics the ideal lowest order resonances for a (possibly long) 
while \cite{BFS86}, but not the higher-order ones whose period (in momentum) 
is not very small compared  
to $n_{ref}$. This is an additional reason preventing experimental 
detection of high-order resonances, no matter how long the 
observation time. Fig.~\ref{fig12} shows a 
simulation for an ensemble of rotors, with a rectangular pulse 
shape of width $\tau _p$. This does not include the smooth switching 
on/off of pulses, as described  e.g. in \cite{SMOR00}; no 
substantial difference is however expected 
in the dynamics on relatively small 
time scales.  
In each kicking interval the rotors 
freely evolved over a physical time $\tau-\tau_p$. During 
the remaining time $\tau _p$ they evolved according to 
the pendulum Hamiltonian $(\hat{\cal N}+\beta)^2/2+
k\cos(\hat{\theta})$. 
The latter evolution was computed by a Trotter-Kato 
discretization of the Floquet operator (equivalent to replacing  the 
pulse by a thick sequence of $\delta-$subkicks). 
\par\vskip 0.2cm \noindent 
(II) The experimental signal-to-noise ratio allows 
only a finite interval of momenta to be observed; 
in \cite{DGOSBG01} this 
border was $n_{cut}= 40$ (data with counteracted gravity). 
 Momenta with $n>n_{cut}$ are not included  in the 
experimental data of \cite{DGOSBG01}.
Due to this fact, the theoretical momentum distributions have to 
be appropriately 
weighed prior to comparison with experimental ones. The crudest 
way is cutting the theoretical distributions  
beyond $n_{cut}$ and renormalizing the probability to $1$.

\par\vskip 0.2cm\noindent

The effect of (I) and (II) on the ideal behaviour discussed in the
previous section is easily understood in qualitative terms. 
We start with the SE-free case.  
The resonant growth of 
energy is stopped as soon as the ballistic peak in the tail approaches 
the closest of the two borders 
that are the effective cut-offs: (I) $n_{ref}$ and (II) $n_{cut}$. 
If this happens earlier than the observation time, then 
the resonant peak is significantly depressed in comparison 
to the ideal case. We shall presently argue that such depression   
is mainly due to the cut-off (II) for the experimental cases 
of  \cite{DGOSBG01,Dar02}. 

In the case of a rectangular pulse, $n_{ref}$ is not a precisely defined 
quantity, due to the slow decay of the Fourier harmonics of the pulse.  
It has to be meant in an effective sense. We hence resort 
to numerical simulations.    
In Fig.~\ref{fig12} numerically computed   
momentum distributions are compared with those obtained in  
the ideal \dk~case; according to such data, the effective 
$n_{ref}$ should be located in the momentum range $70-120$.

The second cut-off $n_{cut}$ is simulated by not counting momenta 
higher than $n_{cut}$ when calculating  
energies, momentum distributions etc. 
(the computational basis of momentum eigenstates is however 
much larger than $n_{cut}$). 
Following experimental parameters  
\cite{DGOSBG01,Dar02} we choose $n_{cut}=40$, the distributions are
renormalized to unity after disregarding states with 
momenta larger/smaller than 
$n=\pm 40$. In Fig.~\ref{fig12a} the effect of this cut-off on the growth 
of the mean energy is shown. In the presence of $n_{cut}$ the deviation 
from the ideal case appears somewhat earlier, as expected from $n_{cut}<
n_{ref}$; moreover, the deviation at $t \gtrsim 20$ is strongly 
enhanced in the presence of $n_{cut}$ (and in the absence of SE). 
As shown in Fig.~\ref{fig13} (a), the momentum distributions 
including both  cut-offs (I) and (II) 
are stable in time, not moving at all in the centre around $n=0$. 
The  slight enhancement at $|n| 
\simeq 15-40$ as compared to the case without cut-offs   
(shown in Fig.~\ref{fig12}) is only due to $n_{ref}$ which to some extent 
acts like a reflecting boundary. 
The ballistic peak, however, which moves in momentum like 
$n \sim \pi kt/2 $ (see discussion after (\ref{decdis})) is lost 
already after about $t\simeq 40/k \simeq 16$ kicks, cf., Fig.~\ref{fig12a}. 
The peak  is then beyond the
cut-off (II).  
The estimated loss after about $16$ kicks is in correspondence with the 
saturation of the mean energy vs. time at quantum resonance  
which has been observed in \cite{DGOCS01,Dar02}
for $t> 15$ in the experimental results, and in the 
theoretical modelling as well (Fig.~(4) in \cite{DGOCS01}). 

\begin{figure} 
\centerline{\epsfig{figure=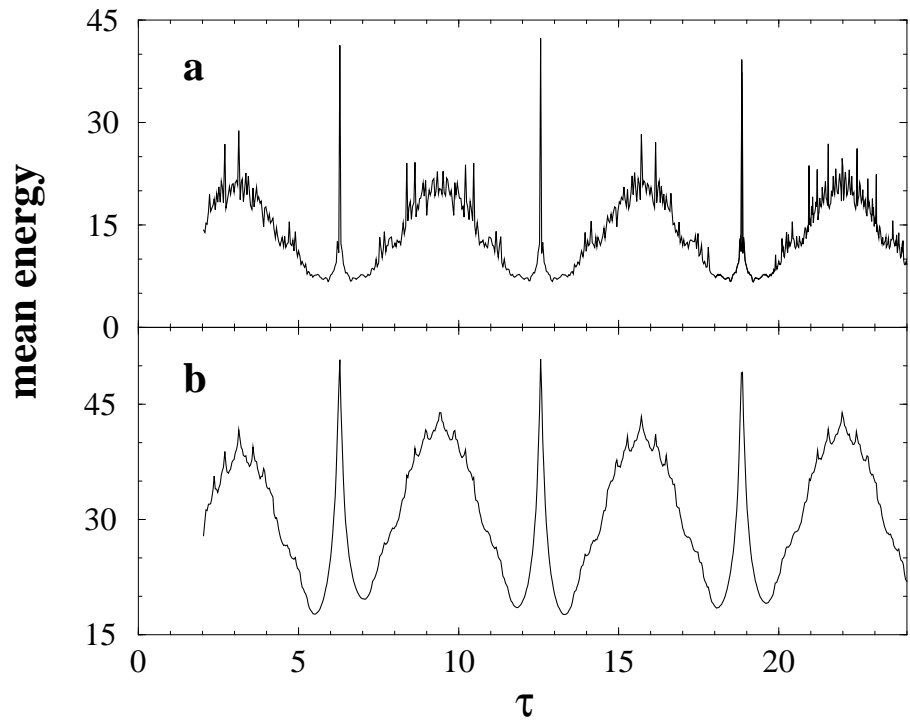,width=10cm,angle=270}} 
\caption{Mean energy after $t=30$ kicks, vs. 
the kicking period $\tau$, for an  ensemble of $10^5$ $\delta-$kicked atoms, 
without momentum cut-offs,   
with the same initial distribution as in Fig.~\ref{fig2} and $k=0.8\pi$.  
(a) no decoherence, (b) added spontaneous emission 
with rate $\pSE=0.2$. Step-size in $\tau$: $\delta \tau \simeq 0.03$, 
avoiding simple rational numbers; same results were obtained for a  
high-resolution random grid in the 
kicking period $\tau$.
}
\label{fig3} 
\end{figure}

\begin{figure} 
\centerline{\epsfig{figure=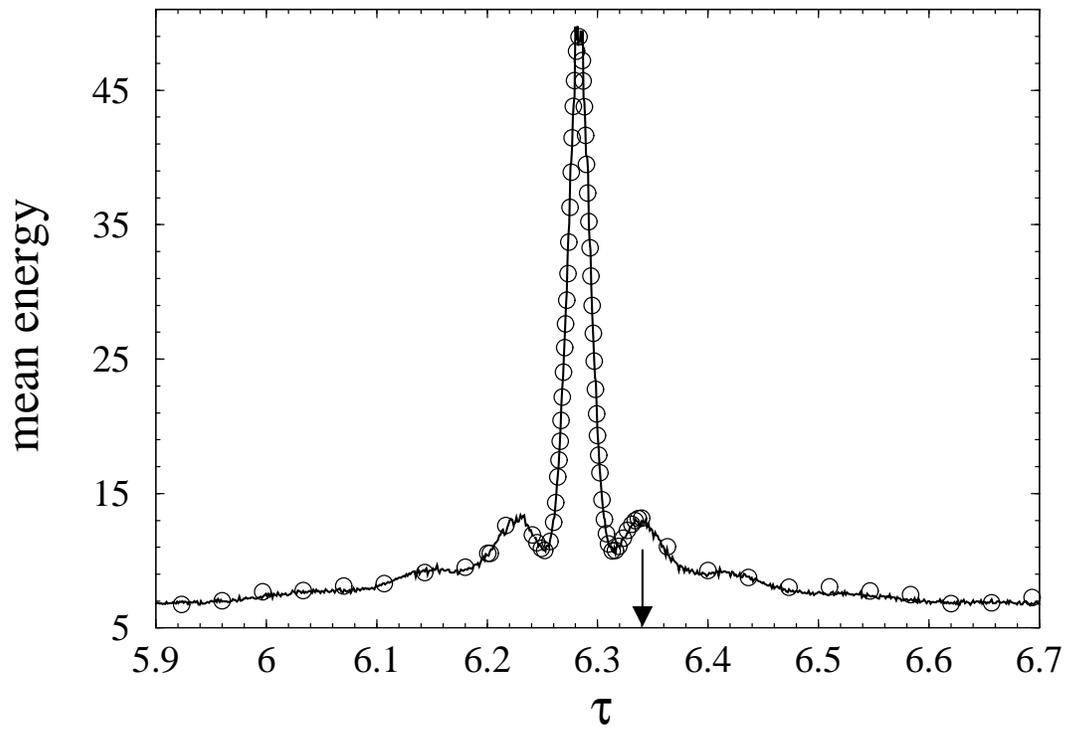,width=10cm,angle=270}} 
\caption{Magnification of Fig.~\ref{fig3}(a) near the 
resonance $\tau=2\pi$. Quantum data taken from 
Fig.~\ref{fig3}(a) (circles) are compared with the mean energies  
of an ensemble of $10^6$ 
$\ep-$classical atoms (solid line) with the same initial momentum  
distribution, evolving under the $\ep-$classical dynamics 
(\ref{clmap}). The value
of $\tau$ corresponding to the small peak on the right of the 
resonant spike is marked by an arrow for reference to Fig.~\ref{fig5}.
} 
\label{fig4} 
\end{figure}

\begin{figure}
\centerline{\epsfig{figure=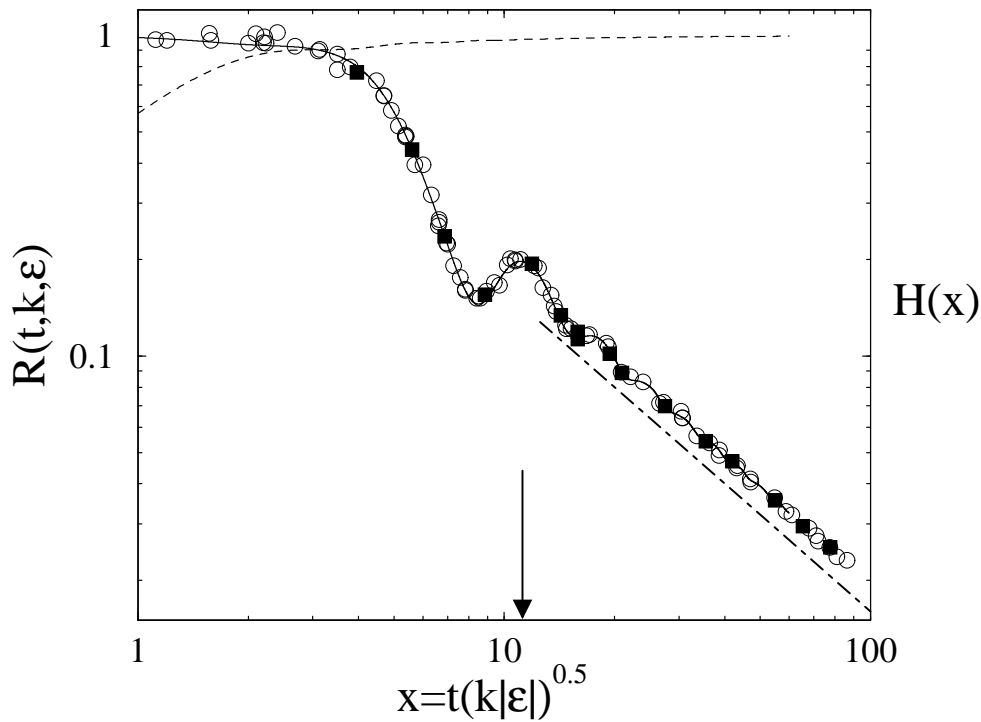,width=10cm,angle=270}} 
\caption{Demonstrating the scaling (\ref{repla0}) of the resonant 
peak, in a right neighbourhood of $\tau=2\pi$. Open circles correspond 
to different values of the parameters $\ep,k,t$, randomly 
generated  in the ranges $1<t<200$, $0.001<\ep <0.1, 
0.1<k<50$ with the constraint $0.01<k\ep <0.2$.
In each case an ensemble  
of $2\times 10^6$ $\ep-$classical rotors was used to numerically 
compute the scaled energy $R(t,k,\ep)$ (\ref{repla0}), with a 
uniform distribution of initial 
momenta  in  $[0,1]$ and a uniform distribution of  
initial $\theta$ in $[0,2\pi)$. 
Full squares present quantum data for $k=0.8 \pi, t=50$ and $t=200$. 
The solid line through the data 
is the scaling function $H(x)$ of (\ref{repla0}) obtained by direct numerical 
computation of the functions  $\Phi_0(x)$ and $G(x)$. The
dashed line represents the function $\Phi _0(x)$; the
dashed-dotted line has slope $-1$ and emphasizes the $x^{-1}$ decay 
described in the text. The arrow marks the value of the scaled 
detuning $x$ which corresponds to the arrow in Fig.~\ref{fig4}. 
}
\label{fig5}
\end{figure}

\begin{figure} 
\centerline{\epsfig{figure=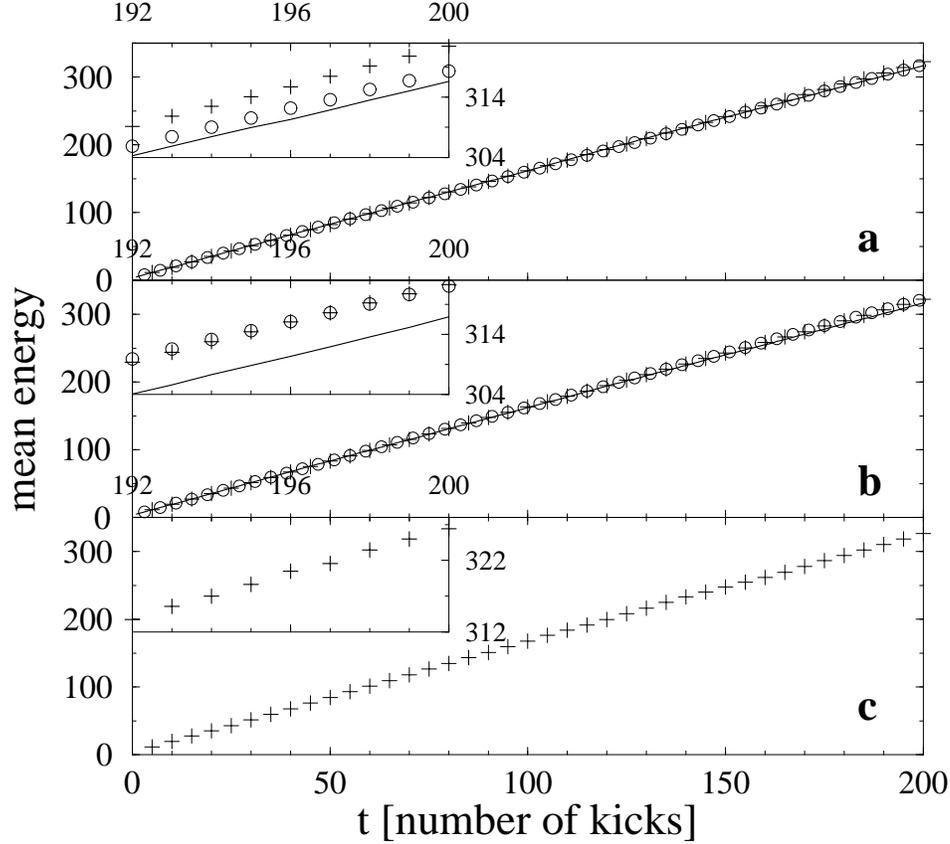,width=11cm,angle=270}} 
\caption{Average energy vs. time $t$ at exact resonance  
$\tau =2\pi$ for the same ensemble of atoms as in Fig.~\ref{fig2}, 
for $k=0.8\pi$, in the presence of SE events 
simulated in different ways as described in the text:  
(a) a random number of SEs  occur 
immediately after kicks, each causing a  momentum change 
$\delta p$ uniformly  distributed in $[-1/2,1/2]$
(type (I) simulation).
(b) SE times  are Poisson-distributed in a window $\tSE=0.067 \tau$, 
with free evolution in-between them;  $\delta p$ is 
distributed as in (a).
(c) SE times as in (a), but $\delta p$ has the parabolic distribution 
(\ref{paraboldensity}) with $k_L/2k_T \simeq 0.476$. 
Rates of spontaneous emission $\pSE =0.05$ (solid),  
$\pSE=0.1$ (circles), $\pSE =0.2$ (plusses).  
The theoretical prediction (see text) for the 
coefficient of linear growth $D_{dec}$ is  
approximately $1.59$,  
whereas the data lead to $D_{dec}\approx 1.58-1.60$ except in 
(b) for $\pSE=0.05$ where it takes the value 
1.55 (strong fluctuations). 
The insets zoom into the region close to $t=200$.
No momentum cutoffs are used.
}
\label{fig6} 
\end{figure}

\begin{figure} 
\centerline{\epsfig{figure=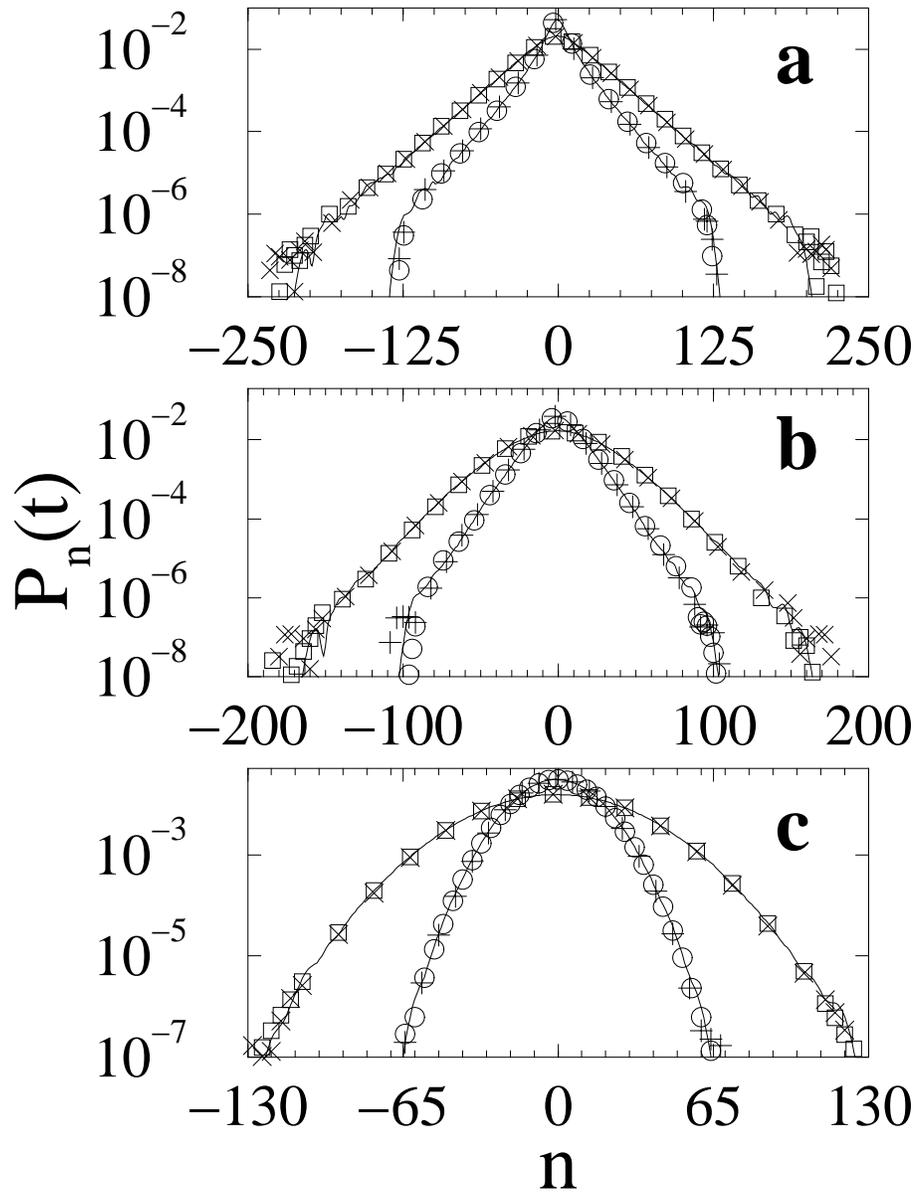,width=12cm}} 
\caption{Evolution in time of coarse-grained momentum distributions 
for the same initial ensemble as in Fig.~\ref{fig2}, and
for $k=0.8$, in the presence of SE.   (a) 
$\pSE =0.1$, (b) $\pSE =0.2$,
(c) $\pSE =0.8$, for $t=50$ and $t=200$. The SE events are simulated 
in different ways. Solid lines were computed like in 
(a) in the previous figure; circles and squares, like in (b); 
plusses and crosses, like in (c) there. 
} 
\label{fig7} 
\end{figure}

\begin{figure}
\centerline{\epsfig{figure=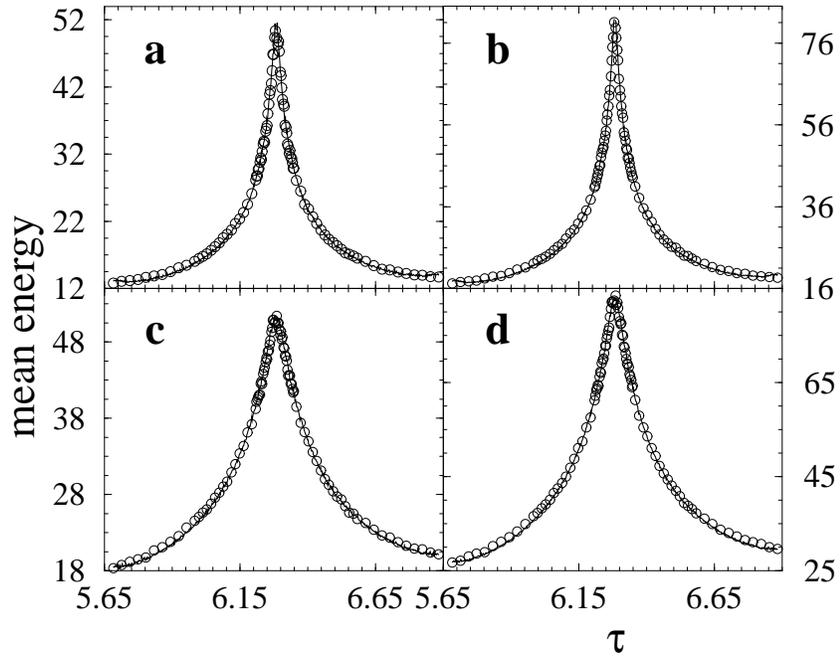,width=10cm,angle=270}} 
\caption{Analogue of Fig.~\ref{fig4}, with the same initial 
ensemble, for $k=0.8\pi$, in the presence 
of SE. Results of full quantum calculations 
(circles) and of $\ep-$classical ones 
(solid lines) in the presence of SE are compared 
near the resonance $\tau=2\pi$, for different times and different rates 
of SE: (a) $\pSE=0.1, t=30$,  (b) $\pSE=0.1, t=50$, (c) $\pSE=0.2, t=30$, 
and  (d) $\pSE=0.2, t=50$. The quantum simulation was type (I), while the  
$\ep-$classical simulations used  the map (\ref{clmapse}).  
}
\label{fig8}
\end{figure}

\begin{figure}
\centerline{\epsfig{figure=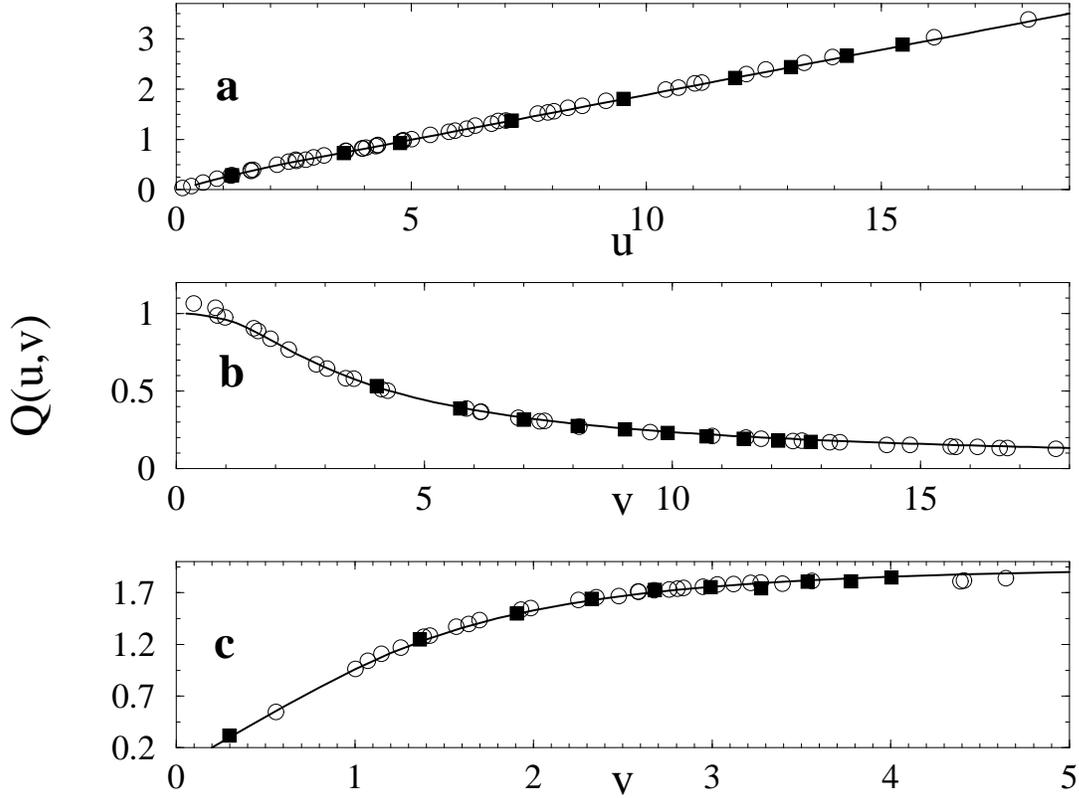,width=11cm,angle=270}} 
\caption{Demonstrating the scaling law (\ref{replace}) 
in a right neighbourhood of $\tau=2\pi$. In (a), (b) the 
quantity on the lhs of eq.~(\ref{scalse}) is plotted vs. one of the
parameters $u=t/t_c$ or  $v=t_c/t_{res}$ while keeping the other fixed: 
(a) $v=2$, (b) $u = 4$. In (c) the ratio $u/v=4$ is fixed.
Open symbols correspond to different values of the parameters 
$t,t_c,k,\ep$, randomly generated  in the ranges $1<t<200$, 
$5<t_c<60$, $0.001<\ep<0.1$, $0.1<k<20$, with the constraints 
$0.001<k\ep<0.2$ and $t_c\sqrt{k\ep}=2$ in (a), $t/t_c=4$ in (b), 
$t=4t_c^2\sqrt{k\ep}$ in (c). In each case an ensemble  
of $2\times 10^{6}$ $\ep-$classical rotors was used, with a 
uniform distribution of initial momenta  in  $[0,1]$ and a 
uniform distribution of initial $\theta$ in $[0,2\pi)$). 
The random momentum shifts at each 
step of the $\ep-$classical evolution (\ref{clmapse}) were generated from 
the  uniform distribution in $[-1/2,1/2]$. 
Full squares represent quantum data for 
$k=0.8\pi$, and $\ep =0.01$ in (a), $\ep =0.05$ in
(c), and $t=50$ and $t=100$ in (b).
The solid lines correspond to the theoretical formula (\ref{scalQ1}).
}
\label{fig9}
\end{figure}

\begin{figure} 
\centerline{\epsfig{figure=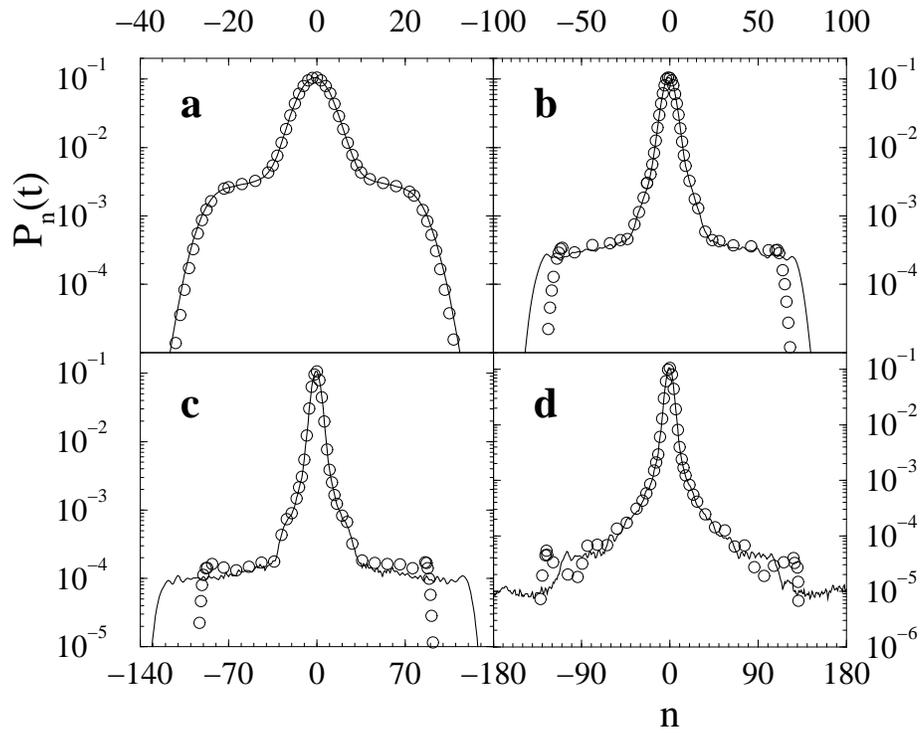,width=10cm,angle=270}} 
\caption{Coarse-grained momentum distributions for 
the same initial ensemble as in Fig.~\ref{fig2} and without SE, 
for $k=0.8\pi$ and $\tau=2\pi$.  The ideal 
case of $\delta-$kicks (solid line) is compared to the case of  
rectangular pulses (open circles). Times are $t=10$ (a), $30$ (b), 
$50$ (c), $200$ (d).
}
\label{fig12} 
\end{figure}

\begin{figure} 
\centerline{\epsfig{figure=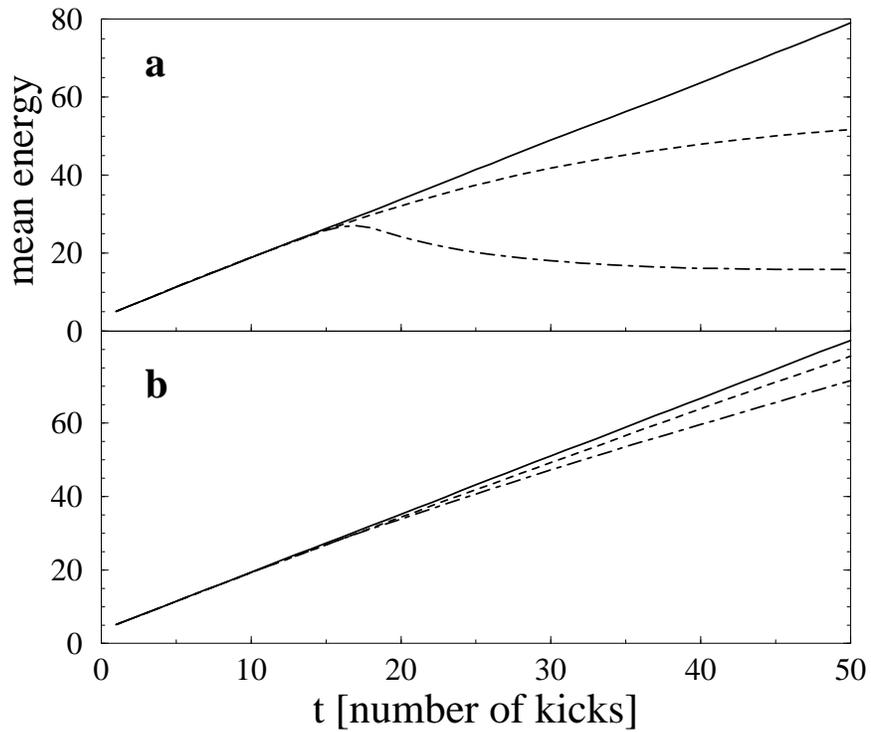,width=10cm,angle=270}}  
\caption{Effect of finite pulse width and of momentum cutoff on the growth 
of the mean energy for the same initial ensemble as in Figs.~\ref{fig2} 
and  \ref{fig12},  
for $k=0.8\pi$ and $\tau=2\pi$, without 
SE (a) and with $\pSE=0.2$ (b). Solid lines are for the ideal $\delta-$kicks 
and no momentum cutoff; dashed lines for rectangular pulses, no cutoff; 
dashed-dotted lines for rectangular pulses and momentum cutoff at 
$n_{cut}=40$. 
In (a) the energy is significantly depressed by the cutoff after  $t>20$,      
not so in (b).}
\label{fig12a} 
\end{figure}

\begin{figure} 
\centerline{\epsfig{figure=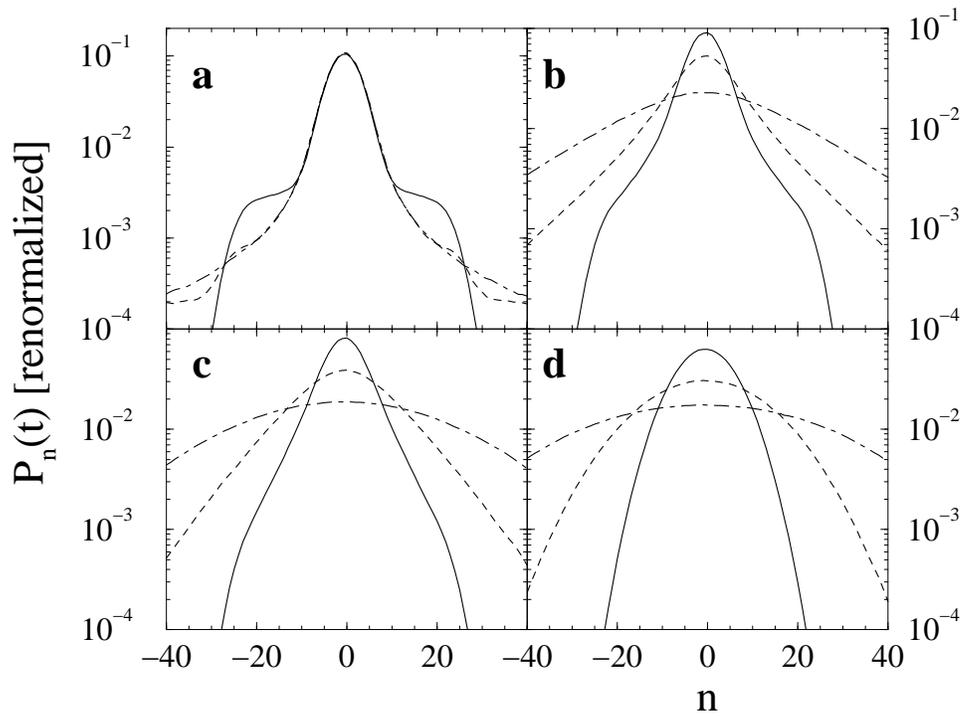,width=10cm,angle=270}} 
\caption{Evolution of coarse-grained momentum distribution 
for the same  ensemble 
as in Figs.~\ref{fig2} and~\ref{fig12}, for 
$\tau =2\pi$ and $k=0.8\pi$, with rectangular pulses and momentum 
cutoff at $n_{cut}=40$. (a) $\pSE=0$, (b) $\pSE=0.1$, (c) $\pSE=0.2$, 
(d) $\pSE=0.8$,  after $t=10$ (solid lines), $t=50$ (dashed),  
$t=200$ (dashed-dotted) 
}
\label{fig13} 
\end{figure}

\begin{figure} 
\centerline{\epsfig{figure=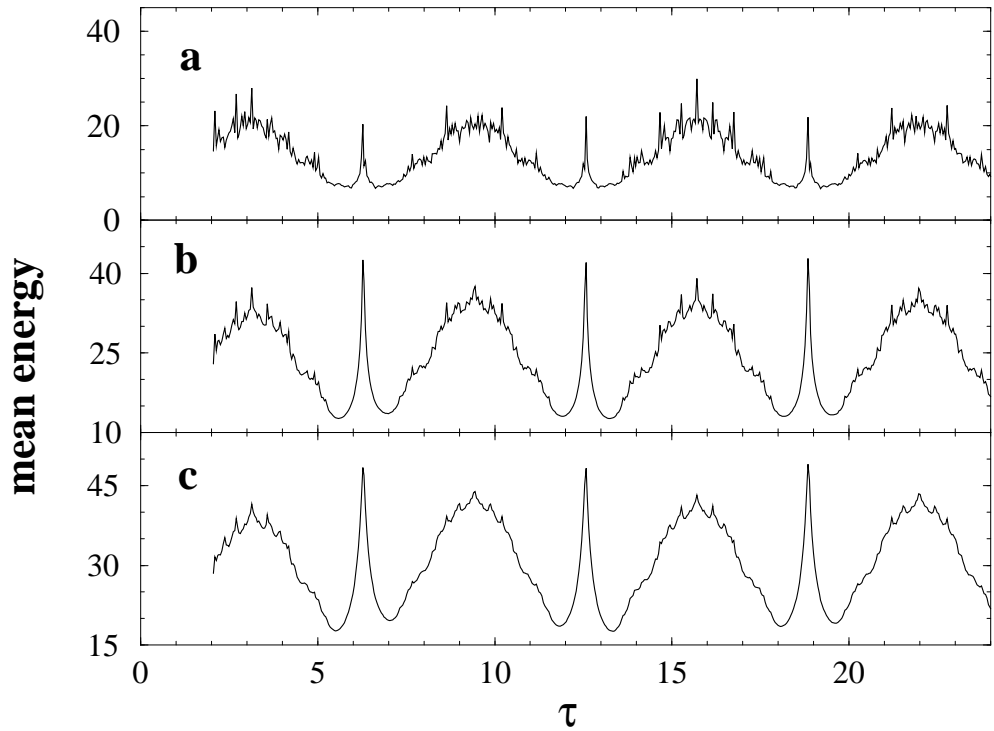,width=10cm,angle=270}} 
\caption{Mean energy as a  
function of the kicking period $\tau$ for the same ensemble of SE 
 and the same initial
distribution of atoms  as in Fig.~\ref{fig3},
after 30 kicks for $k=0.8\pi$.  
(a) no decoherence $\pSE=0$, (b) 
$\pSE=0.1$, (c) $\pSE=0.2$. 
The width of the rectangular pulse is $\tau_p=0.047$ and  $n_{cut}= 40$, 
as in the experiments \cite{DGOCS01,DGOSBG01}. 
The shown range of $\tau$ corresponds in laboratory units 
to $21.2 \ldots 254.7 \mu sec$.
}
\label{fig15} 
\end{figure}

The dependence of the 
mean energy on the kicking period $\tau$, which was shown  in 
Fig.~\ref{fig3} for the ideal case of $\delta-$kicks and no cut-offs, 
is strongly influenced by the cut-offs at exact resonance.
This dependence 
in the absence  of SE, with rectangular pulses and cutoff
at $n_{cut}$ is shown in Fig.~\ref{fig15}(a). By comparing to  
Fig.~\ref{fig3}, we directly see that the only substantial 
difference is at resonant values $\tau =2\pi, 4\pi, 6\pi$:  
cut-offs  lead to lower  resonant peaks. 
When  the cut-off (II) $n_{cut}= 40$ is applied in the ideal case of 
$\delta-$kicks, no differences can be detected from the results plotted  
in Fig.~\ref{fig15}(a), again confirming 
that cut-off (II) is the crucial one. 
The resonance peaks are smaller, because   
the resonant growth of energy stops, 
as soon as the ballistically moving rotors 
hit the boundary (cf., Fig.~\ref{fig12a}). 
Then the mean energy very quickly falls below its ideal value 
(after about 16 
kicks in the plotted case), as can be seen in Fig.~\ref{fig12a}. 

Added SE totally changes this picture. The energy growth is now due 
to the overall broadening of the distribution, and not just to 
the ballistic peaks in the tail, as can be seen comparing the various parts of 
Fig.~\ref{fig7}, where kicks are $\delta-$like, and no cutoff is present. 
The distributions with weak SE 
 are broader in the tails as compared to those 
with strong SE; the latter are however flatter in the centre, which is why 
they have 
roughly the same rms deviation. As already commented, in the SE-free 
case the quasi-momentum is constant in time, and atoms with quasi-momenta 
close to $1/2$ travel faster, thus producing the long 
tails and the ballistic peaks at their edges.  In the presence 
of SE, no atom may persist  a long time in the fast-travelling quasi-momentum 
range, whence it is removed the sooner, the larger $\pSE$. Due to this reason, 
with SE the  cut-offs  are ``felt'' much later 
by the evolving distribution (Fig.~\ref{fig13},~\ref{fig7}).  
Whereas  the cutoff still prevent observation of the 
fastest atoms, they do not significantly affect the growth of energy  
until large times. Even then, 
the momentum distribution normalized within $|n|<n_{cut}$ 
approaches  the flat distribution in $|n|<n_{cut}$ (Fig.~\ref{fig13}(b,c,d)), 
which has a limit value for the second moment significantly higher than 
the SE-free steady state distribution in the presence of the cutoff.

In contrast to the SE-free case, the dependence of the mean 
energy on the kicking period $\tau$ after $30$ kicks 
 is but slightly affected  by the cutoffs 
when SE is present. 
This is shown  in Fig.~\ref{fig15} (b)(c), to be compared to 
Fig.~\ref{fig3}(b). 
In the experiments (Fig.~(2) in \cite{DGOCS01}, Fig.~(6) in \cite{DGOSBG01}), 
the peaks for all cases (a)-(c) are still smaller than in our 
Fig.~\ref{fig15}, which can be explained 
by the extreme sensitivity of the energy at exact resonance 
to all sort of perturbations besides those included in our present 
analysis, and also by difficulties in experimentally tuning to  
the exactly resonant values of $\tau$.  
Additional experimental restrictions, e.g., the 
experienced fluctuations of the potential depth and the 
resulting averaging over 
slightly different experimental realizations 
\cite{DGOSBG01,Dar02,WSDGTPLC02},  
may lead to a further reduction of the peak, especially in the case without  
decoherence which is most sensitive to any kind of disturbance.

\ack 
We thank Andreas Buchleitner, Michael d'Arcy, Gil Summy, and Gregor Veble 
for stimulating comments. S.~W. appreciated detailed discussions 
with Vyacheslav Shatokhin on the atomic decoherence process.
The work was supported by the EU program  
QTRANS RTN1-1999-08400 (S.~W.),  
by the US-Israel Binational Science Foundation (BSF), by the
Minerva Center of Nonlinear Physics of Complex Systems,
and by the fund for Promotion of Research
at the Technion (S.~F.).

{\appendix 

\section{Analysis of the Steady State Distribution (\ref{fourdis}).}
\label{appdis}

\subsection{Proof of estimate (\ref{estdist}).}
\label{appproof}
From (\ref{fourdis}) it follows that 
\begin{equation} 
\label{start}
\sum\limits_{|n|\geq N}M^*_n(t) = 
\frac{1}{2\pi^2}\int_{-\pi}^{\pi}dx\int_{-\pi/2}^{\pi/2}
d\alpha\sum\limits_{|n|\geq N}J^2_n(z)\;\;,\;\;z:=k\sin(x)\csc(\alpha).
\end{equation}
for any positive integer $N$. 
In \ref{bssineq} we show that: 
\begin{equation}
\label{bessineq}
\sum\limits_{|n|\geq N}J^2_n(z)\leq2\left(\frac{ez}{2N}\right)^{2N}\;, 
\end{equation}
We choose $0<\epsilon<\pi$ and we 
use (\ref{bessineq}) to bound the sum in the inner integral in (\ref{start}) 
when  $|\alpha|>\epsilon/2$; otherwise we use the upper  bound $1$.
Noting that $|z|<\pi k/(2\epsilon)$ whenever 
$\pi/2>|\alpha|>\epsilon$, 
\begin{equation}
\label{bessineq1}
\sum\limits_{|n|\geq N}M^*_n(t)\leq\frac{\epsilon}{\pi}+
2\left(\frac{k'e}{2N\epsilon}\right)^{2N}\;.
\end{equation}
where $k':=k\pi/2$. We now minimize the rhs by choosing  
\begin{equation}
\label{optim}
\epsilon=\frac{ek'}{2N}\left(\frac{8\pi N^2}{ek'}\right)^{\frac{1}{2N+1}}
\end{equation}
(which is indeed not larger than $\pi$ whenever 
$N>k\times 1.03\ldots$). Replacing (\ref{optim}) in 
(\ref{bessineq1}) yields  the estimate (\ref{estdist}).

\subsection{Proof of ineq.~(\ref{bessineq}).}
\label{bssineq}

Using the bound 
(\ref{bessbound})  and the power series expansion of 
Bessel functions \cite{AS72}, 
\begin{equation}
\label{bessexp}
\sum\limits_{n=0}^{\infty}J^2_n(z)e^{nr}\leq\sum\limits_{n=0}^{\infty}
\left(\frac{|z|e^{r/2}}{2}\right)^{2n}\frac{1}{(n!)^2}=J_0(i|z|
e^{r/2})\leq e^{|z|e^{r/2}}\;,
\end{equation}
for any real $r$. It follows that:
$$
\sum\limits_{n=N}^{\infty}
J^2_n(z)\leq e^{-Nr}e^{|z|e^{r/2}}\;,
$$
Ineq.~(\ref{bessineq}) follows upon optimizing with respect to $r$.

\subsection{Proof of the asymptotic formula (\ref{decdis}).}

For $z\in[-1,1]$ we  define:
\begin{equation}
\label{funz}
f(z)\;:=\;\frac{1}{2\pi}\int_{-\pi/2}^{\pi/2}d\alpha\;J^2_0(kz\csc(\alpha))\;
\;\mbox{\rm for}\;z\neq 0\;\;;\;\;f(0)=\frac12\;.
\end{equation}
The integrand in (\ref{funz}) is meant $=0$ for $\alpha=0, z\neq 0$. 
Using the integral identity (\ref{bess3}) for  
Bessel functions,  (\ref{fourdis}) may be  
rewritten as: 
\begin{equation} 
\label{fourdisj0} 
M_n^*=\frac{1}{\pi}\int_0^{2\pi}dx\;\cos(2nx) f(\sin(x))\;, 
\end{equation} 
so, for $|n|>0$,  $M^*_n$ is the $2n-$th coefficient in the cosine 
expansion of $f(\sin(x))$. The function  
$f(z)$ is differentiable 
in $[-1,1]\setminus\{0\}$. 
It will be presently shown that
$$
f'(0+)=\lim\limits_{z\to 0+}f'(z)=-\frac{4k}{\pi^2}\neq 0\;.
$$
Since $f(z)$ is an even function, it will follow that its first derivative 
is discontinuous at $z=0$. We choose $\epsilon>0$ and write 
\begin{equation}
\label{inteps}
f(z)\;=\;f_{\epsilon}(z)+g_{\epsilon}(z)\;\;,\;\;f_{\epsilon}(z)\;:=\;
\frac{1}{2\pi}\int_{-\epsilon}^{\epsilon}d\alpha\;J^2_0(kz\csc(\alpha))\;.
\end{equation}
Then $g_{\epsilon}$ is differentiable around $0$, with 
$g'_{\epsilon}(0)=0$. Hence, 
$f'(0+)=f'_{\epsilon}(0+)$. Next we note that if $z>0$, 
$$
f_{\epsilon}'(z)\;=\;\pi^{-1}z^{-1}\int_{-\epsilon}^{\epsilon}
d\alpha\;F(kz\csc(\alpha))\;,
$$
where
$
F(x)\;:=\;xJ_0(x)J_0'(x)$. Noting that 
$$
|\csc(\alpha)-\alpha^{-1}|<c_1\alpha
$$
for $0<|\alpha|<\pi/2$ and some numerical constant $c_1$, 
one easily finds
\begin{eqnarray}
f_{\epsilon}'(z)&=&
\pi^{-1}z^{-1}\int_{-\epsilon}^{\epsilon}d\alpha\;F(kz\alpha^{-1})+O(\epsilon)
\nonumber\\
&=&-2k\pi^{-1}\int_{kz/\epsilon}^{\infty}
du\;u^{-1}J_0(u)J_1(u)\;+\;O(\epsilon)\;,
\end{eqnarray}
where $J'_0(z)=-J_1(z)$ was used.  Letting 
$z\to 0+$ and thereafter $\epsilon\to 0+$ we obtain:
$$
f'(0+)=\lim\limits_{\epsilon\to 0+}f'_{\epsilon}(0+)
=-2k\pi^{-1}\int_0^{\infty}du\;u^{-1}J_0(u)J_1(u)=-\frac{4k}{\pi^2}\;.
$$
The integral was computed by using 
(\ref{bess5}) and then formula 11.4.36 in \cite{AS72}.

Next we recall from (\ref{fourdisj0})
$$
f(\sin x)\;=\;\frac12 M^*_0+\sum\limits_{n=1}^{\infty}
M^*_n\cos(2nx)\;.
$$
According to the above analysis, the derivative of 
this function jumps by $-8k/\pi^2$ at $x=j\pi$ ($j$ any integer). Hence 
the 2nd derivative 
has the singular part $-8k\pi^{-2}\sum_j\delta(x-j\pi)
$, leading to the asymptotic value  
$-16k\pi^{-3}$ for the coefficients in its cosine expansion. 
This yields 
\begin{equation}
\label{asymn}
M^*_n\sim \frac{4k}{\pi^3 n^2}\;\;\mbox{\rm as}\;\; n\to\infty\;.
\end{equation}

\section{Statistics of the process $Z_m$.} 
\label{appa}

\subsection{Independence of the variables $z_j$.}

We show that, for any integers $n,m$, $(m>n)$,  
 the variables $(z_n,...,z_m)$ are independent of the variables 
$(z_0,...,z_k)$ whenever $k\leq n-2$. 
. It suffices to show that $(z_n,...,z_m)$ 
are independent of $(\xise_0,...,\xise_k,\Delta_0,...\Delta_k)$. 
To see this, let $f$ be an arbitrary (Borel) function of 
$m-n+1$ complex variables, and consider:
\begin{equation}
\label{eind}
{\mathcal M}_{k}\equiv
{\cal E}\{f(z_{n},...,z_{m})|\xise_0,...,\xise_k,\Delta_0,...,
\Delta_k\}.
\end{equation}
Looking at (\ref{forthapp}), and recalling that the $\xise_j$ are 
mutually independent,  one notes that $(z_{n},...,z_{m})$ 
depend on $\xise_j,\Delta_j$ $(0\leq j\leq k)$  through the 
factor $\exp(i\sum_0^{k}\xise_j\Delta_j)$, hence only through 
$\sum_0^k\xise_j\Delta_j$ mod$(2\pi)$.  
Therefore, (\ref{eind}) is a function of the variable 
$ \mu_k\equiv\sum_0^k\xise_j\Delta_j$ mod$(2\pi)$  alone: 
${\cal M}_k={\mathcal M}_k(\mu_k)$. Furthermore, 
since $k+1<n$,   
\begin{eqnarray}
{\mathcal M}_{k}(\mu_k)&=&\int dP(\xise_{k+1},\Delta_{k+1})\;
{\mathcal M}_{k+1}(\mu_{k+1})\nonumber\\
&=&
\int dP(\xi_{k+1},\Delta_{k+1})\;
{\mathcal M}_{k+1}(\mu_k+\xise_{k+1}\Delta_{k+1})
\end{eqnarray}
because $\xise_{k+1},\Delta_{k+1}$ are independent of  past variables; 
here $dP(.,.)$ is their  joint distribution.   
Now $\xise_{k+1}$ is independent of the integer $\Delta_{k+1}$, 
and it is uniformly distributed in $(-\pi,\pi)$. Then the integral does 
not depend on  $\mu_k$, so
$$
{\cal E}\{f(z_{n},...,z_{m})|\xise_0,...,\xise_k,\Delta_0,...,
\Delta_k\}={\cal E}\{f(z_{n},...,z_{m})\},
$$
which proves the announced independence property. 

\subsection{Correlations of variables $z_j$.} 

We prove that, for any $j,k\geq 0$ such that $j+k>1$  
\begin{equation}
\label{zmom}
{\cal E}\{z_j z_k^{*}|\Delta\}=\delta_{jk}\Delta_j
\;\;,\;\; 
{\cal E}\{z_jz_k|\Delta\}=0\;.
\end{equation}
whereas 
\begin{eqnarray}
\label{zmom1}
{\cal E}\{z_1z^*_0|\Delta\}&=&{\cal N}(\Delta_0)\equiv
\sum\limits_{j=1}^{\Delta_0}\int\limits_{-\pi}^{\pi}dP(\xi_0) 
e^{ij\xi_0}\;,\nonumber\\
{\cal E}\{|z_0|^2|\Delta\}&=&{\cal M}(\Delta_0)\equiv
\int\limits_{-\pi}^{\pi}dP(\xi_0)
\frac{\sin^2(\xi_0\Delta_0/2)}{\sin^2(\xi_0/2)}
\end{eqnarray}
where
\begin{equation}
dP(\xi_0)=(2\pi\ell)^{-1}d\xi_0
\sum\limits_{j=0}^{l-1}
\;f_0(\beta_j)\;\;,\;\;\beta_j\equiv\frac{\xi_0}{2\pi\ell}+\frac12
+\frac jl\;\mbox{\rm mod}(1)\;.
\end{equation}
is the distribution of $\xi_0$ and $f_0$ is the probability density of the 
initial quasi-momentum (\ref{betadis}). \\
In order to show (\ref{zmom}), we 
denote: 
$$
\alpha_j=e^{i\xise_j}\;\;,\;\;\varphi_j=\sum\limits_{r=0}^{\Delta_j-1}
\alpha_j^r\;
$$
so that:
\begin{equation}
\label{prr}
z_j z_k^{*}=\varphi_j\varphi_k^{*}\prod\limits_{l=0}^{j-1}
\alpha_l^{\Delta_l}\prod\limits_{m=0}^{k-1}\alpha_m^{-\Delta_m}. 
\end{equation}
Let $j>k$, $j+k>1$, Then $j>1$, and the 1st product has the factor 
$\alpha_{j-1}^{\Delta_{j-1}}$. Hence (\ref{prr}) depends 
on $\xise_{j-1}$ via this factor alone (if $j\neq k+1$) 
or via this factor multiplied by $\varphi_{j-1}^{*}$ 
(if $j=k+1$), leading to a factor $e^{i\xi_{j-1}(\Delta_{j-1}-l)}$  with 
$l \leq \Delta_{j-1}-1$. In both cases 
averaging over $\xise_{j-1}$ yields zero because 
$\xise_{j-1}$ is uniformly distributed whenever $j>1$.
 The case $j<k$ is recovered by complex conjugation. 
The 2nd equality in (\ref{zmom}) is straightforward. 
If $j=k$, then from (\ref{forthapp}) it follows that 
\begin{equation}
\label{2mom}
{\cal E}\{|z_j|^2|\Delta\}=
\int\limits_{-\pi}^{\pi}dP(\xi_j)
\frac{\sin^2(\xi_j\Delta_j/2)}{\sin^2(\xi_j/2)}\;. 
\end{equation}
where $dP(\xi_j)$ is the distribution of $\xi_j$. 
For $j>0$, 
$dP(\xi_j)=d\xi_j/(2\pi)$, and the integral is computed according to 
 (\ref{intsinc}).
The 1st equation in 
(\ref{zmom1}) results of a straightforward calculation using 
the definitions (\ref{forthapp}) of $z_0$ and $z_1$.
 
\subsection{Central Limit property.}

The properties of the process $Z_m$ allow to conclude that 
its distribution is asymptotically Gaussian, thanks to known results 
about the Central Limit Theorem for weakly dependent sequences~\cite{Ser68}. 
Isotropy of the limit Gaussian distribution easily follows from 
computing ${\cal E}\{\Rt^2 (\sum_j^N e^{-i\theta}z_j)\}$ (the 
mean square displacement along the direction $\theta$ in $N$ steps).
Using (\ref{zmom}), the result is independent of $\theta$.

\section{Gaussian asymptotics for momentum distributions.}
\label{appb}

Throughout this appendix, the Fourier transform of a function $f$ is 
denoted by $\hat f$. Notations are otherwise identical to those in 
section~\ref{assym}. In particular,
$$
\phi_t(n/{\sqrt t})=\frac{1}{\sqrt{2\pi}}\int du\; {\hat\phi_t}(u) 
e^{inu/{\sqrt t}}\;. 
$$
Replacing this in (\ref{sigcl1}), and using the Bessel function identity 
(\ref{bess4}), 
\begin{eqnarray}
\langle\phi\rangle_t&=&\frac{1}{\sqrt{2\pi}}\int_0^{\infty}
dF_{t}(\rho) 
\int du\;{\hat\phi_t}(u)J_0(2k\rho\sin(u/{2\sqrt t}))\nonumber\\
&=&\frac{2}{\sqrt{2\pi}}\int_0^{\infty}dx\;xe^{-x^2}\int du\;
{\hat\phi_t}(u) J_0(2kx{\sqrt t}\sin(u/{2\sqrt t}))\;
\end{eqnarray}
which in the limit $t\to\infty$ yields
$$
\langle\phi\rangle_{\infty}=\frac{2}{\sqrt{2\pi}}
\int_0^{\infty}dx\;xe^{-x^2}
\int du\;{\hat\phi_{\infty}}(u) J_0(k x u).
$$
Substitution of 
\begin{equation}
\hat{\phi_{\infty}}(u)=\frac{1}{\sqrt{2\pi}}\int dp\; \phi_{\infty}(p) e^{-ipu}
\end{equation}
and of (\ref{bess5}) 
yields the $t\to\infty$ limit:
$$
\langle\phi\rangle_{\infty}
=2\pi^{-1}\int_0^{\infty}dx\; xe^{-x^2}\int_{-kx}^{kx}
dp\;\frac{\phi_{\infty}(p)}{\sqrt{k^2x^2-p^2}}\;.
$$ 
Interchanging integrals, we obtain the same result found  
in the text (equation (\ref{itn})) by different means.

\section{Derivation of eq.~(\ref{scalQ1}).}
\label{poisson}

Let $\Delta_j^1$, $(j=0,1,...)$ 
be real nonnegative, independent random variables 
exponentially distributed with $\En\{\Delta_j^1\}=1$. For $j$ a nonnegative 
integer denote $s_j=\sum_{k=0}^j\Delta_k$, and $s_{-1}\equiv0$. For given 
$u>0$ let $N_u^1\equiv$max$\{j\;:\;s_j<u\}$. We shall compute the expectation 
of the random variable 
\begin{equation}
\label{ap1}
f_u\equiv\sum\limits_{j=0}^{N_u^1}f(\Delta_j^1)+f(u-s_{N_u^1})\;,
\end{equation}
where $f(x)$ is a given nonrandom function; the sum in eq.~(\ref{scalQ}) 
is of this form, with $f(x)=xH(xv)$.
 We write 
$f_u=\sum_{j=0}^{\infty}f_{u,j}$, where
\begin{eqnarray}
\label{ap2}
f_{u,j}&\equiv&\chi(u-s_{j-1})[
\chi(u-s_{j-1}-\Delta_j^1)f(\Delta_j^1)+\nonumber\\
&+&\chi(\Delta_j^1+s_{j-1}-u)f(u-s_{j-1})]\;,
\end{eqnarray}
and $\chi(.)$ is the unit step function. Then, denoting 
$G(x)=\int_0^xdsf(s)e^{-s}$,  
\begin{equation}
\label{ap2a}
\En\{f_{u,j}|s_{j-1}\}=\chi(r)[G(r)+f(r)e^{-r}]\;\;,\;\;
r=u-s_{j-1}\;.
\end{equation}
Therefore,
\begin{eqnarray}
\label{ap3}
\En\{f_u\}&=&\sum\limits_{j=0}^{\infty}
\En\{\En\{f_{u,j}|s_{j-1}\}\}=
G(u)+f(u)e^{-u}+\nonumber\\
&+&\sum\limits_{j=1}^{\infty}
\int_0^udP_j(x)\left[G(u-x)+f(u-x)e^{x-u}\right]\;,
\end{eqnarray}
where $dP_j(x)=dx\;e^{-x}x^{j-1}/(j-1)!$ is the distribution of $s_{j-1}$ 
for $j>0$. Summing over $j$ and replacing the definition of 
$G(x)$ we finally obtain
\begin{equation}
\label{ap5}
\En\{f_u\}=2\int_0^udx\;e^{-x}f(x)+f(u)e^{-u}
+\int_0^udx\;e^{-x}f(x)(u-x)\;.
\end{equation}
Eq.~(\ref{scalQ1}) in the text is obtained on substituting  
$f(x)=xH(xv)$.

\section{Some formulae used in the text.} 
\label{appc}

The following formulas involving Bessel functions 
were used in the text and previous appendices, these are taken from
\cite{AS72} or derived from formulas there (we give the corresponding
numbers in []):
\begin{eqnarray} 
\frac{1}{2\pi}\int_{\alpha}^{\alpha+2\pi}d\theta\; e^{iz\cos
\theta}e^{-in\theta}\; =\; i^nJ_n(z) 
~~~\mbox{[9.1.21]}
\label{bes1}\\
\sum\limits_{n=-\infty}^{\infty}n^2\;J^2_n(x)\; =\;\frac12 x^2 
~~~\mbox{[9.1.76]}
\label{bess1}\\
\int_0^{2\pi}dx\;J_n^2(b\sin(x))\; = \;
\int_0^{2\pi}dx\;\cos(2nx)J_0^2(b\sin(x)) 
~\mbox{[11.4.7/8]}
\label{bess3}\\
\sum \limits_n J^2_n(z)e^{int}\;=\; J_0(2z\sin(t/2))\;  
~~~\mbox{[11.4.8]}
\label{bess4}\\
J_0(z)\;=\;\frac{1}{\pi} \int_{-1}^1 dx \frac{\cos(zx)}{\sqrt{1-x^2}}
~~~\mbox{[9.1.18]}
\label{bess5}\\
|J_n(z)|\;\leq \;\frac{1}{n!}\left\vert\frac{z}{2}\right\vert^{|n|} 
e^{{\cal I}m(z)}\;
~~~\mbox{[9.1.62]}~.
\label{bessbound}
\end{eqnarray}
\\  
The following integral was used 
in calculating  moments of energy:
\begin{equation} 
\label{intsinc} 
\int_0^{2\pi}dx\;\frac{\sin^2(tx)}{\sin^2(x)}=2\pi t\;\;. 
\end{equation}
}

\section*{References} 


\end{document}